\documentclass[sigconf]{acmart}
\usepackage{booktabs} 

\usepackage[english]{babel}
\usepackage{moresize}
\usepackage{amsmath}
\usepackage{algorithmic}
\usepackage{balance}
\usepackage{comment}
\usepackage{paralist}
\usepackage{bm}
\usepackage{pgfplots}
\usetikzlibrary{pgfplots.dateplot}

\usepackage{flushend}
\usepackage[english]{babel}
\usepackage[latin1]{inputenc}
\usepackage{mathrsfs}
\usepackage{graphicx}

\usepackage{amssymb}
\usepackage{amsfonts}
\usepackage{url}
\usepackage{longtable}
\usepackage{rotating}
\usepackage{multirow}
\usepackage{mathrsfs}
\usepackage{subfigure}
\usepackage{enumitem}
\usepackage[linesnumbered,algoruled,boxed,lined]{algorithm2e}
\usepackage{adjustbox}
\usepackage{hyperref}
\usepackage{pgfplots}
\usetikzlibrary{pgfplots.dateplot}
\usepackage{filecontents}
\definecolor{tblue}{RGB}{31,119,180}
\definecolor{torange}{RGB}{255,127,14}
\definecolor{tgreen}{RGB}{44,160,44}
\definecolor{tred}{RGB}{214,39,40}
\definecolor{tpurple}{RGB}{148,103,189}

\newcommand{\hide}[1]{} 

\newcommand{\etal}{\textit{et al}.}

\newcommand{\ie}{\textit{i}.\textit{e}.}
\newcommand{\eg}{\textit{e}.\textit{g}.} 
\newcommand{\wrt}{\textit{w}.\textit{r}.\textit{t}}

\def\model{SimRec}

\setcopyright{none}

\begin{document}
\fancyhead{}


\copyrightyear{2023}
\acmYear{2023} 
\setcopyright{acmlicensed}\acmConference[WWW'23]{Proceedings of the ACM Web Conference 2023}{April 30-May 4, 2023}{Austin, TX, USA}
\acmBooktitle{Proceedings of the ACM Web Conference 2023 (WWW'23), April 30-May 4, 2023, Austin, TX, USA}
\acmPrice{15.00}
\acmDOI{10.1145/3543507.3583196}
\acmISBN{978-1-4503-9416-1/23/04}

\title{Graph-less Collaborative Filtering}


\author{Lianghao Xia}
\affiliation{The University of Hong Kong}
\email{aka\_xia@foxmail.com}

\author{Chao Huang}
\authornote{Chao Huang is the corresponding author.}
\affiliation{The University of Hong Kong}
\email{chaohuang75@gmail.com}

\author{Jiao Shi}
\affiliation{South China University of Technology}
\email{yjjiaoshi@scut.edu.cn}

\author{Yong Xu}
\affiliation{South China University of Technology}
\email{yxu@scut.edu.cn}


\begin{abstract}
Graph neural networks (GNNs) have shown the power in representation learning over graph-structured user-item interaction data for collaborative filtering (CF) task. However, with their inherently recursive message propagation among neighboring nodes, existing GNN-based CF models may generate indistinguishable and inaccurate user (item) representations due to the over-smoothing and noise effect with low-pass Laplacian smoothing operators. In addition, the recursive information propagation with the stacked aggregators in the entire graph structures may result in poor scalability in practical applications. Motivated by these limitations, we propose a simple and effective collaborative filtering model (\model) that marries the power of knowledge distillation and contrastive learning. In \model, adaptive transferring knowledge is enabled between the teacher GNN model and a lightweight student network, to not only preserve the global collaborative signals, but also address the over-smoothing issue with representation recalibration. Empirical results on public datasets show that \model\ archives better efficiency while maintaining superior recommendation performance compared with various strong baselines. Our implementations are publicly available at: \url{https://github.com/HKUDS/SimRec}.
\end{abstract}

\maketitle

\section{Introduction}
\label{sec:intro}

Recent years have witnessed the great success of graph neural network (GNN) in learning latent representations for graph structured data~\cite{zhu2020simple,wu2019simplifying,velivckovic2017graph}. Inspired by such development, many efforts have introduced GNN into Collaborative Filtering (CF) and shown its power in modeling high-order user-item relationships, such as NGCF~\cite{wang2019neural}, LightGCN~\cite{he2020lightgcn}, and GCCF~\cite{chen2020revisiting}. At the core of GNN-based CF models is to utilize a neighborhood aggregation scheme to encode user (item) embeddings via recursively message passing. 

Despite their achieved remarkable performance, we argue that two important limitations exist in current GNN-based CF methods.\\\vspace{-0.12in}

\noindent (i) \textbf{Over-Smoothing and Noise Issues}. The inherent design of GNN may lead to \emph{over-smoothing} issue as the increase of stacked graph layers for embedding propagation~\cite{zhou2020towards,liu2020towards}. The neighborhood aggregator built upon low-pass Laplacian smoothing operation with graph-structured user-item connections, will unavoidably generate indistinguishable user and item representations as the number of layers increases. This results in suboptimal performance as the GNN-based recommenders may fail to capture diverse user preferences. Additionally, various biases of user behavior data widely exist in recommender systems~\cite{chen2021autodebias,zhang2021causal}, such as misclick behaviors, popularity bias. The recursive aggregation schema used in GNN-based models is prone to fusing \emph{noisy} signals, which can hinder the learning of genuine interaction patterns for recommendation. \\\vspace{-0.12in}


\noindent (ii) \textbf{Scalability Limitation with Recursive Expansion}. Although GNN-based recommenders can capture high-order connectivity by stacking multiple propagation layers, the recursive neighbor information aggregation incurs expensive computation in model inference~\cite{yan2020tinygnn,zheng2022bytegnn,gallicchio2020fast}. Therefore, GNN-based CF models with deeper graph neural layers require repeatedly propagate representations among many neighboring nodes, which show poor scalability in practical scenarios, especially for large-scale recommender systems. In light of this, the scalability limitation of current GNN-based methods brings an urge for designing an efficient and effective high-order relation learning paradigm in recommendation, which remains unexplored in existing CF recommendation models.

\begin{figure*}
    \vspace{-0.15in}
    \centering
    \subfigure[Performance v.s. inference time]{
        \includegraphics[height=0.2\textwidth]{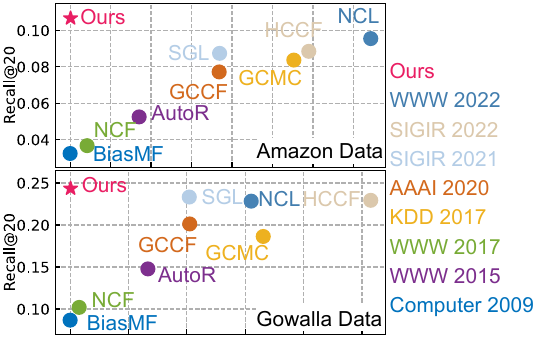}
        \label{fig:intro_time_performance}
    }
    \subfigure[Advantage of our adaptive embedding recalibration]{
        \includegraphics[height=0.2\textwidth]{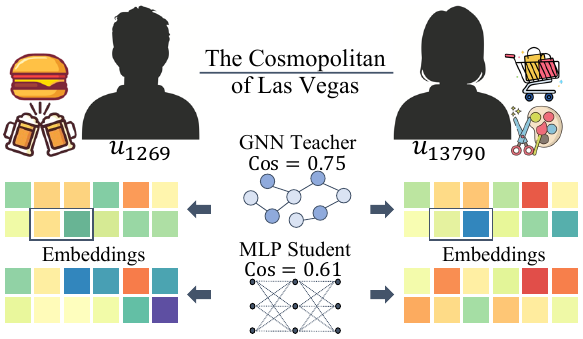}
        \label{fig:intro_case}
    }
    \subfigure[Distribution of learned user embeddings]{
        \includegraphics[height=0.2\textwidth]{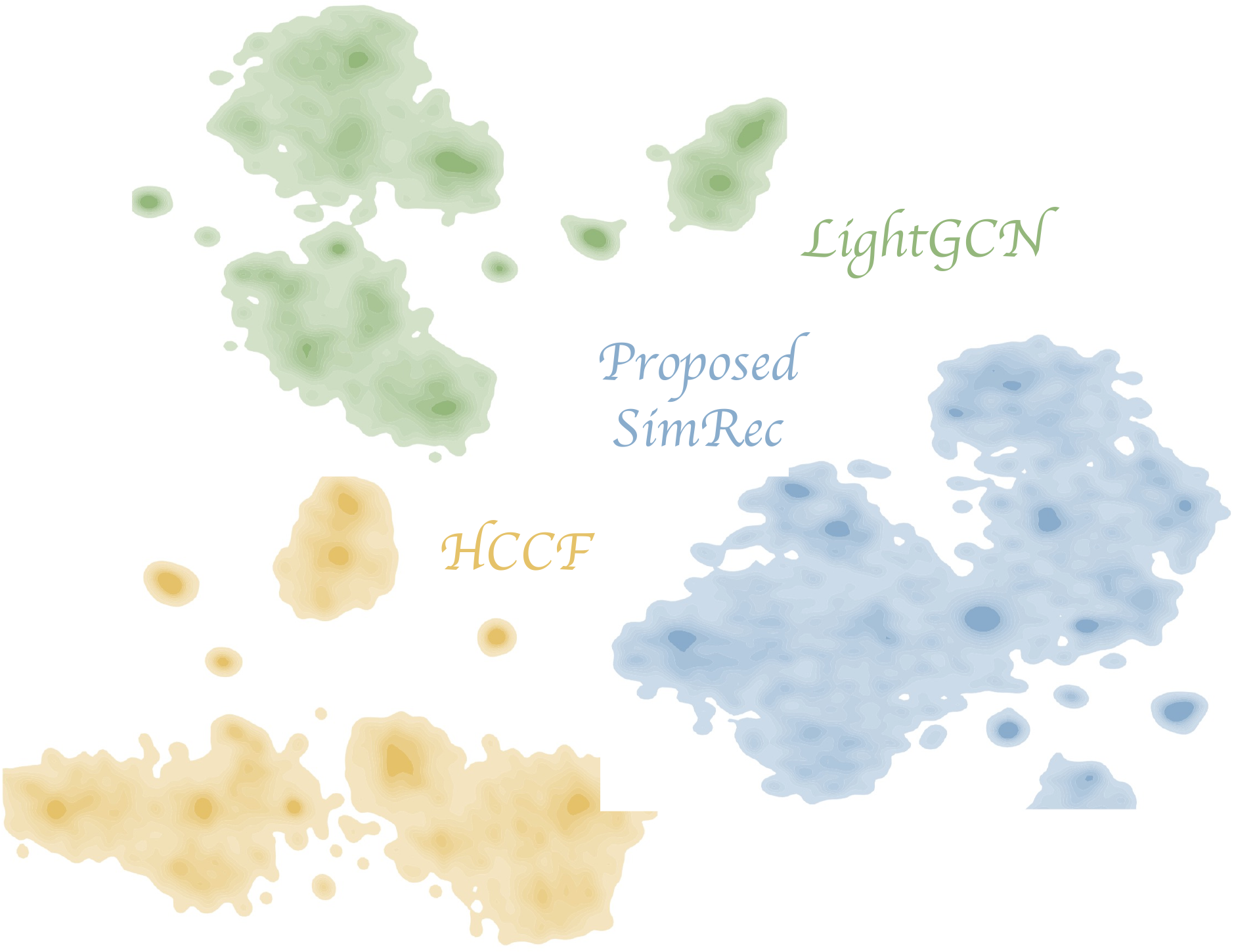}
        \label{fig:intro_dist}
    }
    \vspace{-0.15in}
    \caption{Illustration of motivation and advantages of our \model\ model from different perspectives.}
    \vspace{-0.15in}
    \label{fig:intro}
    \Description{Motivated examples showing the strength of \model\ from three aspects. Firstly, the proposed \model\ achieves best performance with less inference time. Secondly, our \model\ discovers the difference between a noisy user pair. Thirdly, \model\ learns embeddings preserving better user preference uniformly in comparison to baselines.}
\end{figure*}

Having realized the importance of addressing the above challenges, however, it is non-trivial considering the following factors:\vspace{-0.05in}
\begin{itemize}[leftmargin=*]

\item How to well preserve global collaborative signals in an efficient manner for user-item interaction modeling, remains a challenge.

\item How to encode informative representations which are robust to over-smoothing and noise issues while preserving high-order interaction patterns, requiring adaptive knowledge transfer.

\end{itemize}

As shown in Figure~\ref{fig:intro}, we illustrate motivation examples for the model design in our \model\ recommender. To be specific, by comparing our \model\ with various state-of-the-art GNN-based CF methods (\eg, GCCF~\cite{chen2020revisiting}, SGL~\cite{wu2021self}, HCCF~\cite{xia2022hypergraph}) in Figure~\ref{fig:intro_time_performance}, \model\ significantly improves model efficiency, meanwhile maintaining superior recommendation performance. In Figure~\ref{fig:intro_case} of two users with dissimilar interests, we show the advantage of our adaptive contrastive knowledge distillation to recalibrate the similar representations encoded by GNN teacher model into distinguishable embedding space. To reflect the better uniformity preserved in learned embeddings of \model, we visualize the distributions of projected embeddings learned by different methods in Figure~\ref{fig:intro_dist}.\\\vspace{-0.12in}

Inspired by the effectiveness of knowledge distillation (KD) in various domains (\eg, computer vision~\cite{zhang2021data}, text mining~\cite{chen2020distilling}, and graph mining~\cite{zhanggraph}), KD has become an effective solution to transfer knowledge from a large model to a smaller one. In general, KD aims to reach the agreement between the prediction results of a well-trained teacher model and a student model by minimizing their distribution difference. However, collaborative filtering task usually involves highly sparse interaction data, which undermines the capability of knowledge distillation. Specifically, direct distillation from noisy and sparse graph structures, is difficult to advance the performance of original GNN model after being compressed. Fortunately, recent developments of contrastive learning bring new insights in alleviating data sparsity with auxiliary self-supervision signals, this paper explores the possibility of marrying the power of knowledge distillation and contrastive learning to pursue adaptive knowledge transfer with a robust and efficient CF model.

In this work, we propose a novel graph-less collaborative filtering framework, named \model, to improve both the effectiveness and efficiency of recommender without the sophisticated GNN structures. In particular, we propose a bi-level alignment framework to distill knowledge with both prediction-level and embedding-level signals. With such design, the distilled knowledge comes from not only the teacher model's predictions but also the latent high-order collaborative semantics preserved in embeddings. Furthermore, we propose to enhance our knowledge distillation paradigm against the perturbation of over-smoothing and noise effects in GNN teacher model. Towards this end, an adaptive knowledge transfer module is designed with contrastive regularization to capture the diversity of user preference, based on the derived consistency between the supervised CF objective and the augmented SSL task. In our proposed \model\ model, the latent knowledge of GNN-based teacher model will be distilled into a lightweight yet empowered feed-forward network that can jointly capture user-specific preference uniformity and cross-user global collaborative dependencies.

To summarize, our contributions are presented as follows:
\begin{itemize}[leftmargin=*]

\item We propose contrastive knowledge distillation to compress GNN-based CF model into a simple recommender to improve both effectiveness and efficiency. In our adaptive distillation paradigm, an embedding calibration module is designed to enhance KD to preserve useful knowledge and discard the noisy information.


\item Theoretical analysis is provided from two perspectives: i) the benefits of our distillation model in alleviating over-smoothing issue; ii) effectiveness of our distilled self-supervision signals for data augmentation in an adaptive manner.

\item Extensive experiments on public datasets demonstrate that \model\ significantly improves the performance of CF tasks. Additionally, the empirical results show that \model\ gains more efficient embedding encoding over LightGCN on different datasets. 

\end{itemize}

\section{Collaborative Filtering}
\label{sec:model}
In this section, we introduce important notations in collaborative filtering, and recap MLP-based Neural CF and GNN-based CF architectures. In a typical recommendation scenario, there are $I$ users $\{u_1,u_2,...,u_I\}$ and $J$ items $\{v_1, v_2, ..., v_J\}$, indexed by $u_i$ and $v_j$, respectively. An interaction matrix $\textbf{A}\in\mathbb{R}^{I\times J}$ represents the observed interactions between users and items, in which an element $a_{i,j}=1$ if user $u_i$ has adopted item $v_j$, otherwise $a_{i,j}=0$.

Based on the above definitions, a CF-based recommender can be formalized as an inference model that i) $\textbf{Inputs}$ the user-item interaction data $\textbf{A}\in\mathbb{R}^{I\times J}$, models users' interactive patterns based on the input; ii) $\textbf{Outputs}$ interaction prediction results $y_{i,j}$ between the unobserved user-item pair $(u_i, v_j)$. In general, a CF model can be summarized as the following two-stage schema:
\begin{align}
    \label{eq:cf}
    y_{i,j} \leftarrow \textbf{Predict}(\textbf{h}_i, \textbf{h}_j), ~~~~ \textbf{h}_i, \textbf{h}_j \leftarrow \textbf{Embed}(u_i, v_j; \textbf{A})
\end{align}
\noindent The first stage $\textbf{Embed}(\cdot)$ denotes the embedding process which projects user $u_i$ and item $v_j$ into a $d$-dimensional hidden space based on the observed historical interactions $\textbf{A}$. The results of $\textbf{Embed}(\cdot)$ is vectorized representations $\textbf{h}_i, \textbf{h}_j\in\mathbb{R}^d$ for each user $u_i$ and item $v_j$, to preserve user-item interactive patterns. The second stage $\textbf{Predict}(\cdot)$ aims to forecast user-item relations with the prediction score $y_{i,j}\in\mathbb{R}$ using the learned embeddings $\textbf{h}_i, \textbf{h}_j$. 
Based on the above two-stage schema, our proposed method \model\ aims to conduct knowledge distillation from both the embedding and prediction levels for effectively knowledge transferring. \\\vspace{-0.12in}


\noindent\textbf{MLP-based Collaborative Filtering}.
MLP-based neural CF methods~\cite{he2017neural, xue2017deep} are proposed to endow CF with non-linear relation modeling. Due to the simplicity in model architectures, MLP-based CF is highly-efficient and unlikely to learn over-smoothed emebddings like GNNs~\cite{chen2020measuring}. Inspired by the advantages, we adopt MLP as the student model in our contrastive KD framework. In brief, the MLP in \model\ adheres to the two-stage paradigm as follows:
\begin{align}
    y_{i,j} = \textbf{h}_i^\top \textbf{h}_j, ~~~ \textbf{h}_i = \textbf{M-Embed}(\bar{\textbf{h}}_i), ~~~ \textbf{h}_j = \textbf{M-Embed}(\bar{\textbf{h}}_j)
\end{align}
\noindent where $\bar{\textbf{h}}_i, \bar{\textbf{h}}_j\in\mathbb{R}^d$ denote the initial embedding vectors for user $u_i$ and item $v_j$, respectively. $\textbf{M-Embed}(\cdot)$ denotes the MLP-based embedding function. We adopt dot-product for $\textbf{Predict}(\cdot)$, which has been shown to be efficient and effective~\cite{rendle2020neural}. \\\vspace{-0.12in}

\noindent\textbf{GNN-enhanced Collaborative Filtering}.
Most recent CF models apply graph neural information propagation on a bipartite interaction graph $\mathcal{G}=\{\mathcal{U}, \mathcal{V}, \mathcal{E}\}$, to encode users' high-order interactive relations into node embeddings.
Here, $\mathcal{U}=\{u_i\}, \mathcal{V}=\{v_j\}$ denote user and item node sets, respectively. An edge $e_{i,j}\in\mathcal{E}$ exists if and only if $a_{i,j}=1$. Typically, GNN-based CF can be abstracted as:
\begin{align}
    y_{i,j}=\textbf{h}_i^\top\textbf{h}_j,~
    \textbf{H} = \textbf{G-Embed}(\mathcal{G}, \bar{\textbf{H}})=\left(\textbf{Agg}(\textbf{Prop}(\mathcal{G}, \bar{\textbf{H}}))\right)^L
\end{align}
\noindent where $\textbf{H}, \bar{\textbf{H}}\in\mathbb{R}^{(I+J)\times d}$ denote the embedding matrices whose rows are node embedding vectors. $\textbf{G-Embed}(\cdot)$ denotes the GNN-based embedding function which iteratively propagates ($\textbf{Prop}(\cdot)$) and aggregates ($\textbf{Agg}(\cdot)$) the embeddings $\bar{\textbf{H}}$ along the interaction graph $\mathcal{G}$ for $L$ times. Note that though it injects informative structural information, GNNs based on holistic graph modeling and high-order iterations also damage the model scalability and bring the risk of over-smoothing in the collaborative filtering task.

\section{Methodology}
\label{sec:solution}
\begin{figure*}[t]
    \centering
    \includegraphics[width=0.95\textwidth]{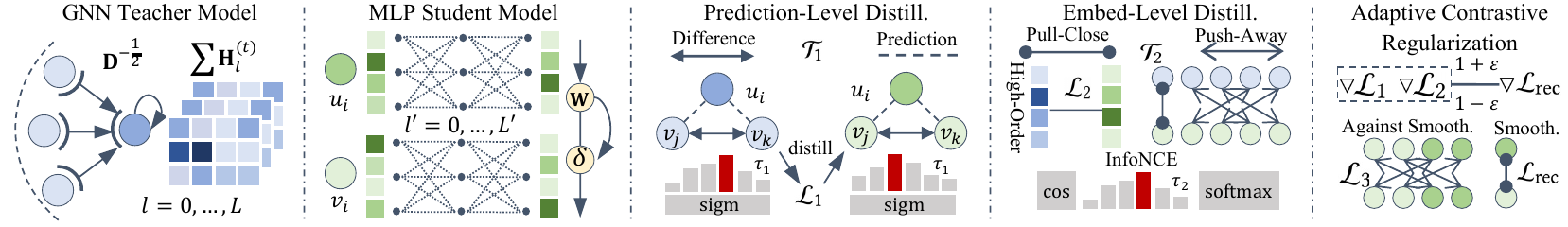}
    \vspace{-0.15in}
    \caption{Model architecture of the proposed \model\ framework.}
    \vspace{-0.1in}
    \label{fig:framework}
    \Description{The architecture of the proposed \model\ model, consisting the GNN teacher, the MLP student, and the bi-level alignment: the prediction-level and the embedding-level knowledge distillation.}
\end{figure*}

In this section, we elaborate the technical details of our proposed \model\ framework, whose workflow is depicted in Figure~\ref{fig:framework}.

\subsection{Contrastive Knowledge Distillation}
For the model design, we are motivated by the advantages of i) GNNs in learning structure-aware node embeddings, and ii) efficient MLPs in preventing over-smoothing issue. Towards this end, we propose to distill knowledge from a GNN-based teacher model to a MLP-based student model. Specifically, the teacher model is a lightweight Graph Convolutional Network (GCN)~\cite{he2020lightgcn,wu2021self,cailightgcl} whose embedding process is shown with the following propagation:
\begin{align}
    \textbf{H}^{(t)}=\sum_{l=0}^L\textbf{H}^{(t)}_{l},~~~ \textbf{H}^{(t)}_{l+1} = \textbf{D}^{-\frac{1}{2}} (\bar{\textbf{A}} + \textbf{I} )\textbf{D}^{-\frac{1}{2}}\cdot \textbf{H}^{(t)}_{l}
\end{align}
where $\textbf{H}^{(t)}\in\mathbb{R}^{(I+J)\times d}$ denotes the embedding matrix given by the teacher model. Index $l$ indicates the number of graph neural iterations (totally $L$ iterations). $\bar{\textbf{A}}\in\mathbb{R}^{(I+J)\times (I+J)}$ denotes the symmetric adjacent matrix for graph $\mathcal{G}$ generated from the interaction matrix $\textbf{A}$ ~\cite{wang2019neural}. $\textbf{I}$ denotes the identity matrix, and $\textbf{D}$ denotes the diagonal degree matrix of $\bar{\textbf{A}}$. The iteration is initialized by $\textbf{H}^{(t)}_{0}=\bar{\textbf{H}}^{(t)}$.
The student model uses a shared MLP network to extract features from the initial embeddings for both users and items. For user $u_i$, the embedding layer is formally presented as follows:
\begin{align}
    \textbf{h}^{(s)}_i=\textbf{FC}^{L'}(\bar{\textbf{h}}^{(s)}_i),~~~~ \textbf{FC}(\bar{\textbf{h}}^{(s)}_i)=\delta(\textbf{W} \bar{\textbf{h}}^{(s)}_i) + \bar{\textbf{h}}^{(s)}_i
\end{align}
\noindent where $\textbf{h}_i^{(s)}, \bar{\textbf{h}}_i^{(s)}\in\mathbb{R}^d$ denote embeddings for $u_i$ given by the student. $\textbf{FC}(\cdot)$ denotes the fully-connected layer. $L'$ is the number of FC layers. An FC layer is configured with one transformation $\textbf{W}\in\mathbb{R}^{d\times d}$. LeakyReLU activation $\delta(\cdot)$, and a residual connection~\cite{he2016deep} are applied. Item-side embedding layer is built analogously.

\subsubsection{\bf Prediction-Level Distillation}
To distill knowledge from the teacher model to the student model, \model\ first follows the paradigm of KL-divergence-based KD~\cite{hinton2015distilling} to align the predictive outputs between the teacher and student models. Inspired by the success of ranking-oriented BPR loss~\cite{rendle2009bpr} in recommender systems, \model\ aligns the two models on the task of ranking user preference. Specifically, in each training step, we randomly sample a batch of triplets $\mathcal{T}_1=\{(u_i,v_j,v_k)\}$, where $u_i, v_j, v_k$ are individually sampled from the holistic user and item set with uniform probability. Then \model\ calculates the preference difference between $(u_i, v_j)$ and $(u_i, v_k)$ for both models, as follows:
\begin{align}
    \label{eq:difference}
    z_{i,j,k} = y_{i,j}-y_{i,k}= \textbf{h}_{i}^{\top}\textbf{h}_{j} - \textbf{h}_{i}^\top\textbf{h}_{k}
\end{align}
where $z_{i,j,k}\in\mathbb{R}$ denotes the difference scores of user preferences for triplet $(u_i, v_j, v_k)$. We denote the score given by the student model as $z_{i,j,k}^{(s)}$ and denote the score given by the teacher model as $z_{i,j,k}^{(t)}$. Then, the prediction-oriented distillation is conducted by minimizing the following loss function:
\begin{align}
    \label{eq:l1}
    \mathcal{L}_1=&\sum\limits_{(u_i,v_j,v_k)\in\mathcal{T}_1}-\left(\bar{z}_{i,j,k}^{(t)} \cdot \log\bar{z}_{i,j,k}^{(s)} + (1 - \bar{z}_{i,j,k}^{(t)}) \cdot \log(1 - \bar{z}_{i,j,k}^{(s)})\right)\nonumber\\
    &\bar{z}_{i,j,k}^{(t)}=\text{sigm}(z_{i,j,k}^{(t)} / \tau_1),~~~~~
    \bar{z}_{i,j,k}^{(s)}=\text{sigm}(z_{i,j,k}^{(s)} / \tau_1)
\end{align}
\noindent where $\bar{z}_{i,j,k}^{(t)}, \bar{z}_{i,j,k}^{(s)}$ are preference differences processed by sigmoid function $\text{sigm}(\cdot)$ with temperature factor $\tau_1$. Here, $z_{i,j,k}^{(t)}$ is given by well-trained teacher model and does not back-propagate gradients. With the help of prediction-oriented distillation $\mathcal{L}_1$, simple MLPs learn to mimic the predictions of advanced GNN models and thus directly generate recommendation results. Through this end-to-end supervision, the parameters of student model are optimized to preserve the knowledge distilled from the teacher model.

It is worth noting that, our prediction-level KD differs from vanilla KD in its training sample enrichment for deep \emph{dark knowledge} learning~\cite{saputra2019distilling, clark2019bam} in CF. Specifically, vanilla KD for multi-class classification~\cite{hinton2015distilling} mines dark knowledge from not only the class with highest score, but also from the ranks for all classes. However, treating CF as multi-classification is problematic, as there are too many classes (items), such that the soft labels easily approach zero and become hard to rank. 
To solve it, our prediction-level KD adopts the pair-wise ranking task instead, and excavates the dark knowledge by distilling from \emph{enriched samples}. Unlike BPR-based model training which pairs each positive item with one negative item, our KD scheme learns from the teacher's predictions on $v_j, v_k$ individually sampled from the holistic item set. Here $v_j, v_k$ are not fixed to be positive or negative. This greatly enriches the training set for our KD and facilitate deeper dark knowledge distillation.

\subsubsection{\bf Embedding-Level Distillation}
Despite the efficacy, the above prediction-level distillation only supervises the model outputs, but ignores the potential difference of embedding distributions between the student and teacher. As both models follow the embedding and prediction schema in Eq~\ref{eq:cf}, we extend the KD paradigm in \model\ with an embedding-level knowledge transferring based on contrastive learning. Specifically, we sample a batch of users and items $\mathcal{T}_2=\{u_i, v_j\}$ from the observed interactions in each training step. Then, we apply the following contrastive loss on the corresponding user/item embeddings:
\begin{align}
    \label{eq:l2}
    \mathcal{L}_2&=
    \sum_{u_i\in\mathcal{T}_2}
    -\log\frac{\exp\left(\cos(\textbf{h}^{(s)}_i, \sum_{l=2}^L\textbf{h}^{(t)}_{i,l})/ \tau_2\right)}{\sum_{u_{i'}\in\mathcal{U}}\exp\left(\cos(\textbf{h}^{(s)}_{i'}, \sum_{l=2}^L\textbf{h}^{(t)}_{i,l}) / \tau_2\right)}\nonumber\\
    & +\sum_{v_j\in\mathcal{T}_2}
    -\log\frac{\exp\left(\cos(\textbf{h}^{(s)}_j, \sum_{l=2}^L\textbf{h}^{(t)}_{j,l})/ \tau_2 \right)}{\sum_{v_{j'}\in\mathcal{V}}\exp\left(\cos(\textbf{h}^{(s)}_{j'}, \sum_{l=2}^L\textbf{h}^{(t)}_{j,l})/ \tau_2\right)}
\end{align}
\noindent where $\cos(\cdot)$ denotes the cosine similarity function. $\tau_2$ represents the temperature hyperparameter. To force the student model to learn more from the high-order patterns which MLP-based CF lacks, here we only use the high-order node embeddings from the teacher. Embeddings of the teacher are well-trained and fixed in parameter optimization. Through directly regularizing the hidden embeddings with this embedding-oriented distillation, \model\ not only further improves the performance of student model, but also greatly accelerates the cross-model distillation, which has been validated in our empirical evaluations.

\subsection{Adaptive Contrastive Regularization}
To prevent transferring over-smoothed signals from the GNN-based teacher to the student model, \model\ proposes to regularize the embedding learning of the student by universally minimizing the node-wise similarity. Specially, \model\ adaptively locates which nodes are more likely being over-smoothed by comparing the gradients of distillation tasks with the main task gradients. In particular, we reuse the sampled users and items $\mathcal{T}_2$ from the embedding-level distillation, and apply the following adaptive contrastive regularization for node embeddings of the student model:
\begin{align}
    \label{eq:l3}
    \mathcal{L}_3=&\sum_{u_i\in\mathcal{T}_2}\varphi(u_i, \mathcal{U}, \omega_i) + \varphi(u_i, \mathcal{V}, \omega_i) + \sum_{v_j\in\mathcal{T}_2}\varphi(v_j, \mathcal{V}, \omega_j)\nonumber\\
    &\varphi(u_i, \mathcal{U},\omega_i) = \omega_i \cdot \log\sum_{u_{i'}\in\mathcal{U}} \exp (\textbf{h}_i^{(s)\top} \textbf{h}_{i'}^{(s)} / \tau_3)
\end{align}
\noindent where the loss $\mathcal{L}_3$ is composed of three terms ($\varphi(\cdot)$) that pushes away the user-user distance, the user-item distance, and the item-item distance, respectively. The first term $\varphi(u_i, \mathcal{U}, \omega_i)$ minimizes the dot-product similarity between the embedding of $u_i$ and the embedding of each user $u_{i'}$ in $\mathcal{U}$, with a weighting factor $\omega_i$. Here, the similarity score is adjusted with the temperature hyperparameter $\tau_3$. The $\phi(\cdot)$ functions for user-item relations and item-item relations work analogously. The weighting factor $\omega_i, \omega_j$ correspond to $u_i, v_j$ respectively, and the weight is calculated as follows:
\begin{align}
    \label{eq:omega}
    \omega_i=\left\{
    \begin{aligned}
        &1-\varepsilon~~~~~\text{if} \bigtriangledown_i^{1,2\top}\bigtriangledown_i^\text{rec}>\bigtriangledown_i^{1^\top}\bigtriangledown_i^2\\
        &1+\varepsilon~~~~~\text{otherwise}
    \end{aligned}
    \right.~~~~~
    &\bigtriangledown_i^{\text{rec}}=\frac{\partial\mathcal{L}_\text{rec}}{\partial\textbf{h}_i^{(s)}}
    \nonumber\\
    \bigtriangledown_i^{1,2}=\frac{\partial(\mathcal{L}_1+\mathcal{L}_2)}{\partial\textbf{h}_i^{(s)}},~~~~~~
    \bigtriangledown_i^{1}=\frac{\partial\mathcal{L}_1}{\partial\textbf{h}_i^{(s)}}, ~~~~~~
    &\bigtriangledown_i^{2}=\frac{\partial\mathcal{L}_2}{\partial\textbf{h}_i^{(s)}}
\end{align}
\noindent where $\omega_i\in\mathbb{R}$ adjusts the weight of contrastive regularization for user $u_i$. In brief, $\omega_i$ has the larger value (\ie, $1+\epsilon$) when the gradients given by distillation tasks (which may over-smooth) contradict to the gradients generated by the main task (which hardly over-smooth). Here, $0<\epsilon<1$ is a hyperparameter. $\bigtriangledown_i\in\mathbb{R}^d$ denotes the gradients for the embedding vector $\textbf{h}_i$ \wrt, different optimization tasks. For example, $\bigtriangledown_i^{1,2}$ denotes the compound gradients of the two distillation task objectives $\mathcal{L}_1$ and $\mathcal{L}_2$. $\bigtriangledown_i^\text{rec}$ denotes gradient of the recommendation task, which is independent to the GNN-based teacher and thus has no risk of over-smoothing. The task $\mathcal{L}_\text{rec}$ will be elaborated later. The similarity between the gradients is estimated using dot-product. When the similarity between the distillation tasks and the recommendation task, is larger than the similarity between two distillation tasks, we can assume that the difference in optimization between the distillation and the recommendation is small enough to weaken the regularization.

\subsection{Parameter Learning of \model}
Following the training paradigm of knowledge distillation, our \model\ first trains the GNN-based teacher model until convergence. In each step, \model\ samples a batch of triplets $\mathcal{T}_\text{bpr} = \{(u_i,v_j,v_k)|a_{i,j}=1, a_{i,k}=0\}$ where $u_i$ denotes anchor user. $v_j$ and $v_k$ denotes positive item and negative item, respectively. The BPR loss function~\cite{rendle2009bpr} is applied on the sampled data as follows:
\begin{align}
    \label{eq:teacher_bpr}
    \mathcal{L}^{(t)}=-\sum_{(u_i,v_j,v_k)\in\mathcal{T}_\text{bpr}}\log\text{sigm}(y^{(t)}_{i,j}-y^{(t)}_{i,k}) + \lambda^{(t)}\|\bar{\textbf{H}}^{(t)}\|_\text{F}^2
\end{align}
\noindent where the last term denotes the weight-decay regularization with weight $\lambda^{(t)}$ for preventing over-fitting.

Then, \model\ conducts joint training to optimize the parameters of the MLP-based student, during which the structure-aware node representations are distilled from advanced GNNs to over-smoothness-resistant MLPs. The training process is elaborated in~\ref{sec:learn_alg}. Strengthened by the two distillation tasks and the regularization terms, the overall optimization objective is presented:
\begin{align}
    \mathcal{L}^{(s)}&=\mathcal{L}_\text{rec} + \lambda_1\cdot \mathcal{L}_1 + \lambda_2\cdot \mathcal{L}_2 + \lambda_3\cdot \mathcal{L}_3 + \lambda_4\cdot\mathcal{L}_4\nonumber\\
    \mathcal{L}_\text{rec} &= -\sum\nolimits_{(u_i,v_j)\in\mathcal{T}_2} y_{i,j},~~~~~~~
    \mathcal{L}_4 = \|\bar{\textbf{H}}^{(s)}\|_\text{F}^2
\end{align}
\noindent where $\lambda_1, \lambda_2, \lambda_3, \lambda_4$ are weights for different optimization terms. $\mathcal{T}_2$ denotes the aforementioned set containing user-item pairs sampled from the observed interactions $\mathcal{E}$. As the contrastive regularization $\mathcal{L}_3$ minimizes the similarity between negative user-item pairs, the recommendation objective $\mathcal{L}_\text{rec}$ only maximizes the similarity between positive user-item pairs. $\mathcal{L}_4$ denotes the weight-decay regularization for the MLP neural network.

\subsection{Further Discussion of \model}
\subsubsection{\bf Adaptive High-Order Smoothing via KD}
\label{sec:highorder_smoothing}
An important strength of GNN-based CF lies in its ability to smooth user/item embeddings using their high-order neighbors. Through derivation, we show our method is able to perform the high-order smoothing in an adaptive manner. Detailed derivations are presented in~\ref{sec:embed_analysis}. In brief, for our light-weight GCN teacher, the embedding parameters $\bar{\textbf{h}}_i^{(t)}, \bar{\textbf{h}}_j^{(t)}$ of two nodes $n_i, n_j$ (either user or item nodes) are smoothed using each other, when minimizing the following terms from the BPR loss $\mathcal{L}^{(t)}$ in Eq~\ref{eq:teacher_bpr}:
\begin{align}
    \label{eq:gcn_gradient}
    \frac{\partial\mathcal{L}^{(t)}_{i,j}}{\partial\bar{\textbf{h}}_i^{(t)}} &= \sum_{v_k}- \sigma \cdot
    \Big(\sum_{\mathcal{P}_{i,j}^{2L}} \prod_{(n_a,n_b)\in\mathcal{P}_{i,j}^{2L}} \frac{1}{\sqrt{d_a d_b}}\Big)\cdot
    \frac{\partial\bar{\textbf{h}}_i^{(t)\top}\bar{\textbf{h}}_j^{(t)}}{\partial \bar{\textbf{h}}_i^{(t)}}
\end{align}
\noindent where $\mathcal{L}_{i,j}^{(t)}$ denotes the terms that pull close the embeddings of $n_i$ and $n_j$ in loss $\mathcal{L}^{(t)}$. $\sigma\in(0,1)$ is a BPR-relevant factor. $\mathcal{P}_{i,j}^{2L}$ represents a possible path between $n_i$ and $n_j$ with maximum length $2L$. $d_a, d_b$ denotes the node degrees of $n_a$ and $n_b$, respectively. Eq~\ref{eq:gcn_gradient} reveals that GCNs smooth embeddings for high-order nodes with weighted gradients. The weights (\ie, the bracketed part) encode how closely nodes are connected via multi-hop graph walks.
Similarly, we analyze the gradients from our prediction-level KD $\mathcal{L}_1$ over embedding parameters $\textbf{h}_i^{(s)}$, as follows:
\begin{align}
    \label{eq:pd_gradient}
    \frac{\partial\mathcal{L}_{i,j}^{(1)}}{\partial{\textbf{h}}_i^{(s)}} = \sum_{v_k} -\frac{1}{\tau_1} \cdot (\bar{z}_{i,j,k}^{(t)} - \bar{z}_{i,j,k}^{(s)}) \cdot \frac{\partial\textbf{h}_i^{(s)\top} \textbf{h}_j^{(s)}}{\partial\textbf{h}_i^{(s)}}
\end{align}
\noindent where $\mathcal{L}_{i,j}^{(1)}$ denotes the part from $\mathcal{L}_1$ that maximizes the similarity between the embeddings of $n_i$ and $n_j$. Eq~\ref{eq:pd_gradient} shows that, by utilizing the prediction-level KD, our MLP-based student can also be supercharged with high-order embedding smoothing without the cumbersome holistic-graph information propagation. Furthermore, the weights for different node pairs (\ie, the bracketed part) are derived from a well-trained GCN model, instead of depending on handcrafted heuristic manners as in Eq~\ref{eq:gcn_gradient}. This makes our KD framework robust to the noise of observed graph structures.


\subsubsection{\bf Enriched Supervision Augmentation via KD}
Recent works~\cite{wu2021self, lin2022improving, xia2022hypergraph} propose to address the noise and the sparsity problems of CF by providing self-supervision signals using contrastive learning (CL) techniques. We show that our KD approach can provide even more additional supervisions. Specifically, we list the pull-close gradients from both InfoNCE-based CL loss, and our KD loss $\mathcal{L}_1$, \wrt, a single node embedding $\textbf{h}_i$, as follows:
\begin{align}
    \label{eq:overall_loss}
    \frac{\partial\mathcal{L}_\text{CL}}{\partial\textbf{h}_i} &= -\frac{1}{\tau} \cdot
    \frac{\partial \textbf{h}_i^{'\top} \textbf{h}''_i / (\|\textbf{h}'_i\|_2 \|\textbf{h}''_i\|_2)}{\partial \textbf{h}_i}\nonumber\\
    \frac{\partial\mathcal{L}_{1}}{\partial{\textbf{h}}_i^{(s)}} &= \sum_{v_j}\frac{\partial\mathcal{L}_{i,j}^{(1)}}{\partial{\textbf{h}}_i^{(s)}} = \sum_{v_j, v_k} -w_{i,j,k} \cdot \frac{\partial\textbf{h}_i^{(s)\top} \textbf{h}_j^{(s)}}{\partial\textbf{h}_i}
\end{align}
\noindent where $w_{i,j,k}$ represents the factors for simplicity. Shown by the second equation, our KD generates $|\{v_j,v_k|(u_i,v_j,v_k)\in\mathcal{T}_2\}|$ pull-close optimization terms for each node $u_i$, while CL method only generates one training sample. This evidently shows that our KD-based scheme can enrich the supplementary supervision signals, even without the data augmentation in CL~\cite{wu2021self}.

\subsubsection{\bf Complexity Analysis}
We analyze the complexity of \model\ to answer the following questions: i) How do GCNs compared to MLPs in efficiency? ii) How is the efficiency of our KD paradigm compared to state-of-the-art methods? Detailed analysis is presented in~\ref{sec:complexity_analysis}. In concise, the computational complexity of the MLP network in \model\ is $\mathcal{O}(|\mathcal{T}_2|\times L' \times d^2)$, and the complexity of the GCN teacher is $\mathcal{O}(|\mathcal{E}|\times {L}\times d)$. The MLP student is more efficient to the GNN teacher. For the second question, the supplementary losses in \model\ takes $\mathcal{O}(|\mathcal{T}_2|\times (I+J)\times d)$ complexity, which is comparable to existing SSL collaborative filtering methods.

\section{Evaluation}
\label{sec:eval}

We conduct experiments from different aspects to validate the efficacy of the propose \model\ framework. The implementation details for our \model\ and the baseline methods are presented in~\ref{sec:implement}. Our experiments aim to answer the following research questions:
\begin{itemize}[leftmargin=*]
    \item \textbf{RQ1}: How does the proposed \model\ perform on different experimental datasets in comparison to state-of-the-art baselines?
    \item \textbf{RQ2}: How does different sub-modules of the proposed \model\ framework contribute to the overall performance?
    \item \textbf{RQ3}: How scalabile is \model\ in handling large-scale data?
    \item \textbf{RQ4}: How does the model performance vary when tuning important hyperparameters of the proposed \model\ model?
    \item \textbf{RQ5}: How can our \model\ model address the over-smoothing issue compared with GNN-based recommendation methods?
\end{itemize}

\subsection{Experimental Settings}
\subsubsection{\bf Experimental Datasets}
\begin{table}[t]
    \centering
    \caption{Statistics of the experimental datasets.}
    \label{tab:datasets}
    \small
    \vspace{-0.18in}
    \begin{tabular}{ccccc}
        \toprule
        Dataset & \# Users & \# Items & \# Interactions & Interaction Density \\
        \midrule
        Gowalla & 25,557 & 19,747 & 294,983 & $5.85\times 10^{-4}$\\
        Yelp & 42,712 & 26,822 & 182,357 & $1.59\times 10^{-4}$\\
        Amazon & 76,469 & 83,761 & 966,680 & $1.51\times 10^{-4}$\\
        \bottomrule
    \end{tabular}
    \vspace{-0.15in}
    \Description{A table showing the statistics of the Gowalla data (25557 users, 19747 items, 294983 interactions), the Yelp data (42712 users, 26822 items, 182357 interactions), and the Amazon data (76469 users, 83761 items, 966680 interactions).}
\end{table}

\begin{table*}[h]
\vspace{-0.1in}
\caption{Performance comparison on Gowalla, Yelp, and Amazon datasets in terms of \textit{Recall} and \textit{NDCG}.}
\vspace{-0.15in}
\centering
\footnotesize
\setlength{\tabcolsep}{1.2mm}
\begin{tabular}{|c|c|c|c|c|c|c|c|c|c|c|c|c|c|c|c|c|c|l|}
\hline
Data & Metric & BiasMF & NCF & AutoR & PinSage & STGCN & GCMC & NGCF & GCCF & LightGCN & DGCF & SLRec & NCL & SGL & HCCF & \emph{\model} & p-val.\\
\hline
\multirow{4}{*}{Gowalla}
&Recall@20 & 0.0867 & 0.1019 & 0.1477 & 0.1235 & 0.1574 & 0.1863 & 0.1757 & 0.2012 & 0.2230 & 0.2055 & 0.2001 & 0.2283 & 0.2332 & 0.2293 & \textbf{0.2434} & $2.1e^{-8}$\\
&NDCG@20 & 0.0579 & 0.0674 & 0.0690 & 0.0809 & 0.1042 & 0.1151 & 0.1135 & 0.1282 & 0.1433 & 0.1312 & 0.1298 & 0.1478 & 0.1509 & 0.1482 & \textbf{0.1592} & $1.2e^{-9}$\\
\cline{2-18}
&Recall@40 & 0.1269 & 0.1563 & 0.2511 & 0.1882 & 0.2318 & 0.2627 & 0.2586 & 0.2903 & 0.3181 & 0.2929 & 0.2863 & 0.3232 & 0.3251 & 0.3258 & \textbf{0.3399} & $2.4e^{-8}$\\
&NDCG@40 & 0.0695 & 0.0833 & 0.0985 & 0.0994 & 0.1252 & 0.1390 & 0.1367 & 0.1532 & 0.1670 & 0.1555 & 0.1540 & 0.1745 & 0.1780 & 0.1751 & \textbf{0.1865} & $1.7e^{-9}$\\
\hline

\multirow{4}{*}{Yelp}
&Recall@20 & 0.0198 & 0.0304 & 0.0491 & 0.0510 & 0.0562 & 0.0584 & 0.0681 & 0.0742 & 0.0761 & 0.0700 & 0.0665 & 0.0806 & 0.0803 & 0.0789 & \textbf{0.0823} & $3.7e^{-4}$\\
&NDCG@20 & 0.0094 & 0.0143 & 0.0222 & 0.0245 & 0.0282 & 0.0280 & 0.0336 & 0.0365 & 0.0373 & 0.0347 & 0.0327 & 0.0402 & 0.0398 & 0.0391 & \textbf{0.0414} & $3.8e^{-5}$\\
\cline{2-18}
&Recall@40 & 0.0307 & 0.0487 & 0.0692 & 0.0743 & 0.0856 & 0.0891 & 0.1019 & 0.1151 & 0.1175 & 0.1072 & 0.1032 & 0.1230 & 0.1226 & 0.1210 & \textbf{0.1251} & $4.8e^{-3}$\\
&NDCG@40 & 0.0120 & 0.0187 & 0.0268 & 0.0315 & 0.0355 & 0.0360 & 0.0419 & 0.0466 & 0.0474 & 0.0437 & 0.0418 & 0.0505 & 0.0502 & 0.0492 & \textbf{0.0519} & $2.4e^{-4}$\\
\hline

\multirow{4}{*}{Amazon}
&Recall@20 & 0.0324 & 0.0367 & 0.0525 & 0.0486 & 0.0583 & 0.0837 & 0.0551 & 0.0772 & 0.0868 & 0.0617 & 0.0742 & 0.0955 & 0.0874 & 0.0885 & \textbf{0.1067} & $1.1e^{-10}$\\
&NDCG@20 & 0.0211 & 0.0234 & 0.0318 & 0.0317 & 0.0377 & 0.0579 & 0.0353 & 0.0501 & 0.0571 & 0.0372 & 0.0480 & 0.0623 & 0.5690 & 0.0578 & \textbf{0.0734} & $7.0e^{-12}$\\
\cline{2-18}
&Recall@40 & 0.0578 & 0.0600 & 0.0826 & 0.0773 & 0.0908 & 0.1196 & 0.0876 & 0.1175 & 0.1285 &0.0912 & 0.1123 & 0.1409 & 0.1312 & 0.1335 & \textbf{0.1535} & $6.6e^{-10}$\\
&NDCG@40 & 0.0293 & 0.0306 & 0.0415 & 0.0402 & 0.0478 & 0.0692 & 0.0454 & 0.0625 & 0.0697 & 0.0468 & 0.0598 & 0.0764 & 0.0704 & 0.0716 & \textbf{0.0879} & $2.0e^{-12}$\\
\hline
\end{tabular}
\vspace{-0.1in}
\label{tab:overall_performance}
\Description{A table presenting the evaluated performance of the proposed \model\ model and the baselines, in which \model\ significantly outperforms the baseline methods.}
\end{table*}


Three benchmark datasets collected from real-world online services are used to evaluate the performance of \model. Data statistics are shown in Table~\ref{tab:datasets}. We split the interaction data into training set, validation set and test set with 70\%:5\%:25\%. Details of the experimental datasets are:
\begin{itemize}[leftmargin=*]
    \item \textbf{Gowalla}: This dataset is collected from Gowalla, including user check-in records at geographical locations, from Jan to Jun, 2010.
    \item \textbf{Yelp}: This dataset contains users' ratings on venues, collected from Yelp platform. The time range is from Jan to Jun, 2018.
    \item \textbf{Amazon}: This dataset is composed of users' rating behaviors over books collected from Amazon platform, during 2013.
\end{itemize}

\vspace{-0.1in}
\subsubsection{\bf Evaluation Protocols}
Following previous works on CF recommenders~\cite{wang2019neural, xia2022self}, we conduct all-rank evaluation, in which positive items from test set are ranked with all un-interacted items for each user. The widely-used \emph{Recall@N} and \emph{NDCG@N} metrics~\cite{wu2021self,2021knowledge} are used adopted for evaluation, where $N=20$ by default.

\vspace{-0.05in}
\subsubsection{\bf Baseline Models}
We compare \model\ with the following 14 baselines from 4 research lines for comprehensive validation.
\\\noindent\textbf{Traditional Collaborative Filtering Technique:}
\begin{itemize}[leftmargin=*]
    \item \textbf{BiasMF}~\cite{koren2009matrix}: It is a classic matrix factorization approach that combines user/item biases with learnable embedding vectors.
\end{itemize}
\textbf{Non-GNN Neural Collaborative Filtering}:
\begin{itemize}[leftmargin=*]
    \item \textbf{NCF}~\cite{he2017neural}: It is an early study of deep learning CF model that enhances the user-item interaction modeling with MLP networks.
    \item \textbf{AutoR}~\cite{sedhain2015autorec}: This method applies a three-layer autoencoder with fully-connected layers to encode user interaction vectors.
\end{itemize}
\textbf{Graph Neural Architectures for Collaborative Filtering}:
\begin{itemize}[leftmargin=*]
    \item \textbf{PinSage}~\cite{ying2018graph}: This method combines random walk with graph convolutions for web-scale graph in recommendation.
    \item \textbf{STGCN}~\cite{zhang2019star}: This method augments GCN with autoencoding sub-networks on hidden features for better inductive inference.
    \item \textbf{GCMC}~\cite{berg2017graph}: This is a representative work to introduce graph convolutional operations into the matrix completion task.
    \item \textbf{NGCF}~\cite{wang2019neural}: It is a GNN-based CF method which conducts graph convolutions on the user-item interaction graph for embeddings.
    \item \textbf{GCCF}~\cite{chen2020revisiting} and \textbf{LightGCN}\cite{he2020lightgcn}: These two methods propose to simplify conventional GCN structures by removing transformations and activations for improving performance.
\end{itemize}
\textbf{Disentangled GNN-based Collaborative Filtering}:
\begin{itemize}[leftmargin=*]
    \item \textbf{DGCF}\cite{wang2020disentangled}: This method disentangles user-item interactions into multiple hidden factors in the graph message passing process.
\end{itemize}
\textbf{Self-Supervised Learning Approaches for Recommendation}:
\begin{itemize}[leftmargin=*]
    \item \textbf{SLRec}~\cite{yao2021self}: This method applies contrastive learning to recommendation models with feature-level data augmentations.
    \item \textbf{NCL}~\cite{lin2022improving}: This approach enhances self-supervised graph CF models with enriched neighbor-wise contrastive learning.
    \item \textbf{SGL}~\cite{wu2021self}: It conducts various types of graph augmentations and feature augmentations with graph contrastive learning for CF.
    \item \textbf{HCCF}~\cite{xia2022hypergraph}: This method augments GNN-based CF with a global hypergraph GNN and conducts cross-view contrastive learning.
\end{itemize}

\subsection{Overall Performance Comparison (RQ1)}

The overall performance of \model\ and the baselines are shown in Table~\ref{tab:overall_performance}. From the results we have the following observations: \vspace{-0.05in}
\begin{itemize}[leftmargin=*]
    \item Our \model\ consistently achieves best performance compared to baselines methods. Also, we re-train \model\ and the best-performed baselines (\ie, SGL and NCL) for 5 times to calculate $p$-values. The experimental results validate the significance of the improvement by \model. Compared to the state-of-the-art GNN methods, the MLP-based inference model of our graph-less \model\ generates more accurate recommendation results, due to its adaptive contrastive knowledge distillation. Specifically, the dual-level KD in \model\ enables enriched and adaptive high-order smoothing, which not only distills the accurate dark knowledge in the well-trained GNN teacher, but also avoids being affected by the over-smoothing signals. Furthermore, the adaptive contrastive regularization automatically alleviates the over-smoothing effects, which further boosts the performance. \\\vspace{-0.12in}
    
    \item While the self-supervised learning schema greatly improves the performance of GNN-based CF, our graph-less \model\ model still significantly outperforms the SSL-enhanced graph models. We attribute the performance deficiency to the inherent incapability of existing SSL frameworks in filtering over-smoothing signals. For example, SGL augments model training by introducing random noises, which may even aggravate the inaccuracy in node embeddings when the noises are magnified through high-order graph propagation. As for NCL and HCCF, they seek to connect nodes based on global semantic relatedness, which may even over-smooth nodes distant from each other in the original graph. In comparison, our graph-less \model\ model abandons GNN architectures in the inference model, which fundamentally minimizes the possibility of over-smoothed node embeddings. Furthermore, our KD paradigm avoids distilling over-smoothed embeddings via the adaptive contrastive regularization. \\\vspace{-0.12in}
    
    \item We observe that non-GNN CF models (\ie, NCF and AutoR) present very bad performance, event though they have similar MLP-based network architectures as the inference model in \model. This sheds light on the deficiency of MLPs in modeling high-order graph connectivity into user/item embeddings. While sharing similar MLP structures, our \model\ is additionally supervised by knowledge distilled from advanced GNN models. This not only improves the optimization for MLP networks, but also makes it possible to adaptively filter the over-smoothing signals in parameter learning. The huge performance gap between NCF/AutoR and our \model\ strongly shows the effectiveness of our contrastive knowledge distillation.
\end{itemize}

\begin{table}[t]
    \caption{Ablation study on key components of \model.}
    \vspace{-0.15in}
    \centering
    \footnotesize
    \begin{tabular}{c|c|cc|cc|cc}
        \hline
        \multicolumn{2}{c|}{Data}& \multicolumn{2}{c|}{Gowalla} & \multicolumn{2}{c|}{Yelp} & \multicolumn{2}{c}{Amazon}\\
        \hline
        \multicolumn{2}{c|}{Variant} & Recall & NDCG & Recall & NDCG & Recall & NDCG\\
        \hline
        \hline
        \multicolumn{2}{c|}{-$\mathcal{L}_1$} & 0.2180 & 0.1415 & 0.0756 & 0.0377 & 0.1012 & 0.0692\\
        \hline
        \multirow{3}{*}{-$\mathcal{L}_2$} & User & 0.2292 & 0.1493 & 0.0806 & 0.0405 & 0.0998 & 0.0667 \\
        & Item & 0.2266 & 0.1477 & 0.0808 & 0.0406 & 0.0974 & 0.0649 \\
        & Both & 0.2222 & 0.1451 & 0.0787 & 0.0399 & 0.0938 & 0.0626 \\
        \hline
        \multirow{4}{*}{-$\mathcal{L}_3$} & U-I & 0.2330 & 0.1496 & 0.0814 & 0.0410 & 0.0939 & 0.0607 \\
        & U-U & 0.2349 & 0.1512 & 0.0811 & 0.0407 & 0.0965 & 0.0634 \\
        & I-I & 0.2331 & 0.1514 & 0.0813 & 0.0409 & 0.1009 & 0.0674\\
        & All & 0.2282 & 0.1480 & 0.0810 & 0.0407 & 0.0933 & 0.0605\\
        \hline
        \hline
        \multicolumn{2}{c|}{\emph{\model}} & \textbf{0.2434} & \textbf{0.1592} & \textbf{0.0823} & \textbf{0.0414} & \textbf{0.1067} & \textbf{0.0734}\\
        \hline
    \end{tabular}
    \vspace{-0.1in}
    \label{tab:module_ablation}
    \Description{A table presenting the results of module ablation study. The results are divided into three parts: loss $\mathcal{L}_1$ for the prediction-level distillation, loss $\mathcal{L}_2$ for the embedding level distillation, and loss $\mathcal{L}_3$ for the contrastive regularization. All ablated variants performs worse than the proposed \model.}
\end{table}

\vspace{-0.1in}
\subsection{Model Ablation Study (RQ2)}
\begin{figure}[t]
    \centering
    \includegraphics[width=0.43\columnwidth]{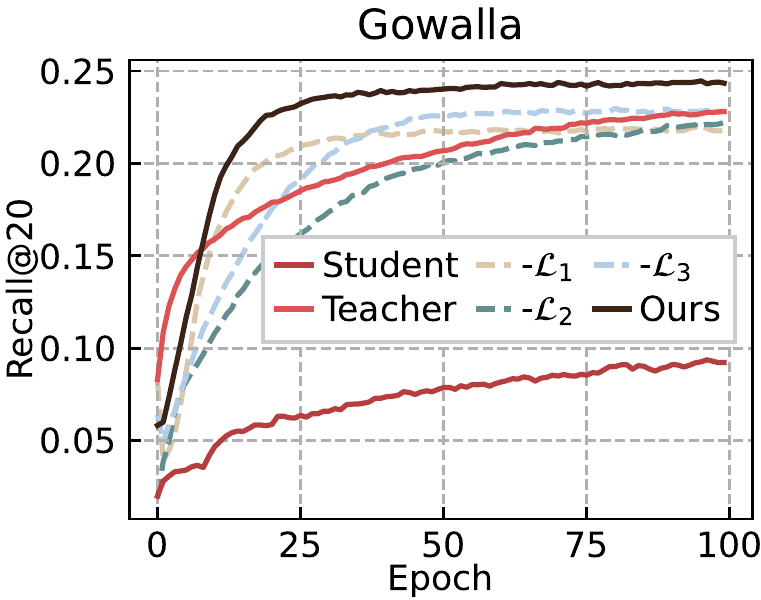}\quad
    \includegraphics[width=0.43\columnwidth]{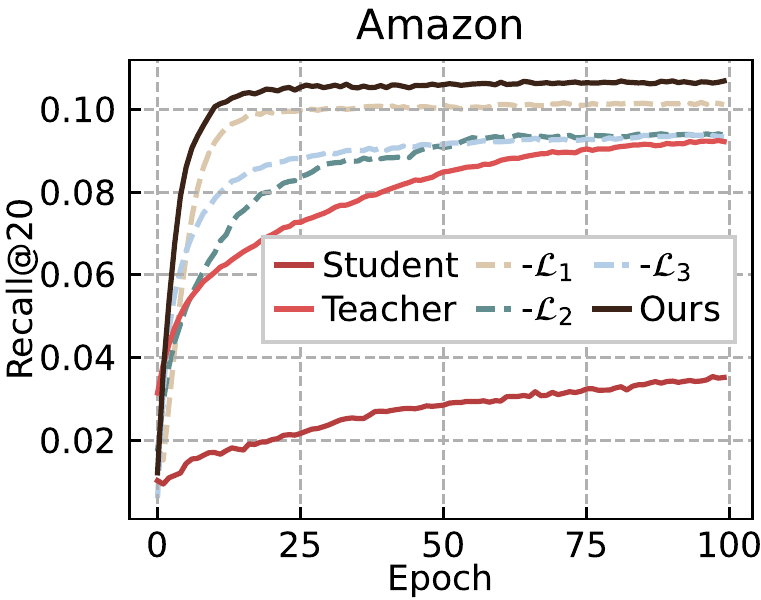}
    \vspace{-0.12in}
    \caption{Test performance in each epoch for ablated models.}
    \vspace{-0.1in}
    \label{fig:ablation_lines}
    \Description{A line figure showing the performance with respect to epochs for \model\ and some representative baselines. The figure shows that \model\ converges faster while training.}
\end{figure}

We validate the effectiveness of the applied sub-modules in \model\ by ablating each module separately. The evaluated performance is shown in Table~\ref{tab:module_ablation}. We also show the performance change \wrt, training epochs in Figure~\ref{fig:ablation_lines}. We have the following observations:
\begin{itemize}[leftmargin=*]
    \item \textbf{Effect of Prediction-Level Distillation}: Our prediction-level distillation (\ie, $\mathcal{L}_1$) excavates deep dark knowledge in the teacher using the pair-wise ranking task with enriched KD samples. The variant -$\mathcal{L}_1$ removes this module, which leads to performance degradation on Gowalla and Yelp data. The results validate the effectiveness of learning from the predictive outputs of teacher model using our distillation loss $\mathcal{L}_1$.\\\vspace{-0.12in}
    \item \textbf{Effect of Embedding-Level Distillation}: We then test the effect of embedding-level KD with the variant -$\mathcal{L}_2$ by removing $\mathcal{L}_2$ on user/item embeddings. In some cases the alignment between users and the alignment between items have different effect on the performance. What's more, the results reveal not only the contribution of $\mathcal{L}_2$ to the final performance, but also its prominent accelerating effect in model training shown in Fig~\ref{fig:ablation_lines}. \\\vspace{-0.12in}
    \item \textbf{Effect of Contrastive Regularization}: We ablate \model\ without the contrastive regularization in variant -$\mathcal{L}_3$. The regularization for user-item, user-user, and item-item relatedness are individually ablated. We observe the importance of $\mathcal{L}_3$ for the superior performance, especially on Amazon data. We ascribe this to the larger scale of Amazon data which makes it more likely to over-smooth with irrelevant high-order neighbors. The incorporation of $\mathcal{L}_3$ can cancel out over-smoothing signals.\\\vspace{-0.12in}
    \item \textbf{Comparison to Student and Teacher Models}: From the learning curves in Fig~\ref{fig:ablation_lines}, we can observe the great performance gap between simple MLP student and advanced GNN teacher. The three augmented tasks greatly minimizes this gap by effectively distilling useful knowledge. Additionally, the distillation tasks accelerate the training to surpass the original teacher model.
\end{itemize}

\begin{table}[t]
    \centering
    \footnotesize
    \setlength{\tabcolsep}{1.4mm}
    \caption{Model performance and per-epoch model inference time of representative methods on large-scale Tmall dataset.}
    \label{tab:scalability}
    \vspace{-0.1in}
    \begin{tabular}{ccccccc}
        \hline
        Metric & \# Edges & DGCF & SGL & HCCF & NCL & \emph{\model}\\
        \hline
        \hline
        \multirow{2}{*}{R@20} & 1.6M & 0.0221 & 0.0258 & 0.0272 & 0.0286 & \multirow{2}{*}{\textbf{0.0308}}\\
        & 2.9M & 0.0253 & 0.0278 & 0.0283 & 0.0294 & \\
        \hline
        \multirow{2}{*}{N@20} & 1.6M & 0.0258 & 0.0296 & 0.0309 & 0.0337 & \multirow{2}{*}{\textbf{0.0366}}\\
        & 2.9M & 0.0279 & 0.0311 & 0.0319 & 0.0334 & \\
        \hline
        \multirow{2}{*}{Time} & 1.6M & 7190.2s & 1331.8s & 1342.5s & 1392.2s & \multirow{2}{*}{\textbf{785.1s}}\\
        & 2.9M & 11431.8s & 1456.3s & 1530.8s & 1693.8s & \\
        \hline
    \end{tabular}
    \vspace{-0.12in}
    \Description{A table showing the performance and the inference time of \model\ and baselines on the large-scale Tmall dataset. \model\ outperforms the baselines and consumes the least time for inference.}
\end{table}

\vspace{-0.1in}
\subsection{Model Scalability Study (RQ3)}
To validate the efficiency of our \model\ in handling large-scale real-world data, we compare \model\ with the best performed baselines on a e-commerce data collected from Tmall platform. The dataset contains around 40 million records of user clicks. To successfully run on this dataset, GNN-based methods have to sample subgraphs for information propagation. In contrast, graph sampling is not required by the MLP-based inference model of our \model. The performance and the inference time are shown in Table~\ref{tab:scalability}, where we run the baselines using graph sampling strategy~\cite{hu2020heterogeneous} with two scales (\ie, subgraphs contain 1.6M edges and 2.9M edges, respectively). We have mainly two key observations shown as follows:
\begin{itemize}[leftmargin=*]
    \item \textbf{More Accurate Recommendations}: \model\ achieves better recommendation performance in terms of Recall and NDCG. This reflects the higher probability of over-smoothing on the large but sparse interaction graph. Our \model\ avoids this problem without explicit graph message passing. Instead, informative knowledge is distilled from GNNs for model compression.
    \item \textbf{Much Higher Efficiency}: \model\ greatly reduces the inference time on the large Tmall data. \textit{Firstly}, the embedding process of our MLP predictor is agnostic to the holistic interaction graph, thus the large-scale graph does not increase much overhead for embedding processing. No graph sampling is required in comparison to GNNs. \textit{Secondly}, \model\ infers user-item relations based on simple MLPs. The computational costs of fully-connected layers in MLPs are much lower than the cost of GNNs.
\end{itemize}

\begin{figure}[t]
    \centering
    \includegraphics[width=0.3\columnwidth]{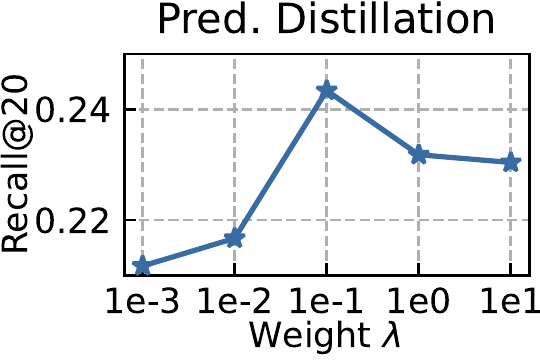}\quad
    \includegraphics[width=0.3\columnwidth]{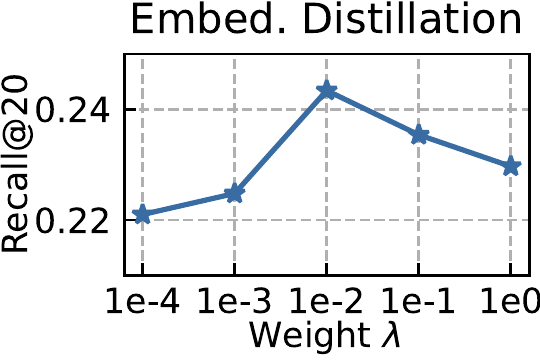}\quad
    \includegraphics[width=0.3\columnwidth]{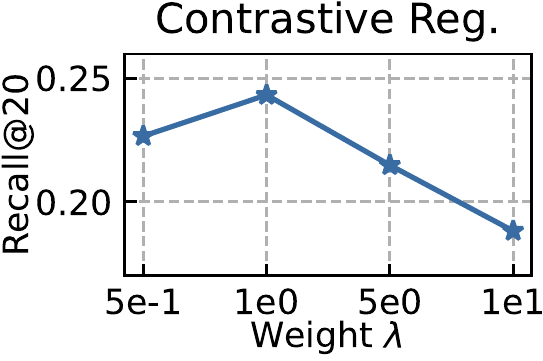}\\
    \includegraphics[width=0.3\columnwidth]{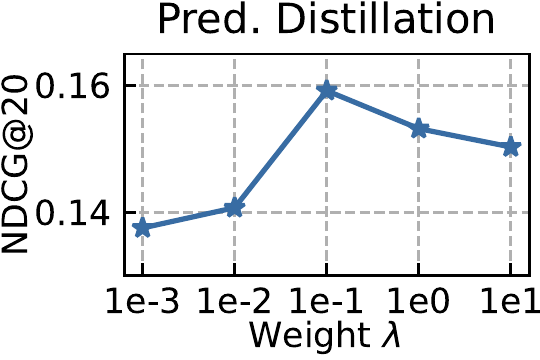}\quad
    \includegraphics[width=0.3\columnwidth]{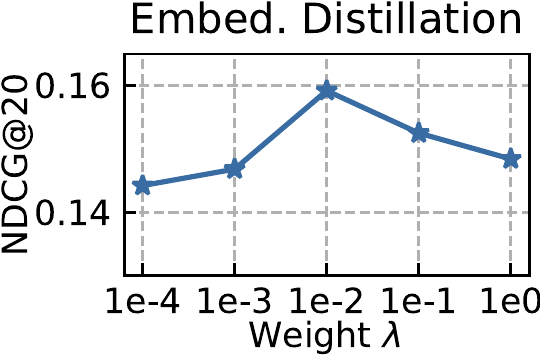}\quad
    \includegraphics[width=0.3\columnwidth]{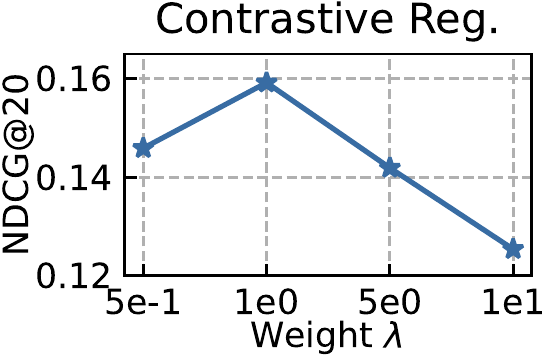}\\
    \vspace{-0.12in}
    \caption{Hyperparameter study for our \model\ model on Gowalla dataset, in terms of \emph{Recall@20} and \emph{NDCG@20}.}
    \vspace{-0.1in}
    \label{fig:hyper2d}
    \Description{A line figure showing the performance change with respect to the weight of the prediction-level distillation, the embedding-level distillation, and the contrastive regularization.}
\end{figure}

\begin{figure}[t]
    \centering
    \subfigure[Pred. Distillation]{
        \includegraphics[width=0.3\columnwidth]{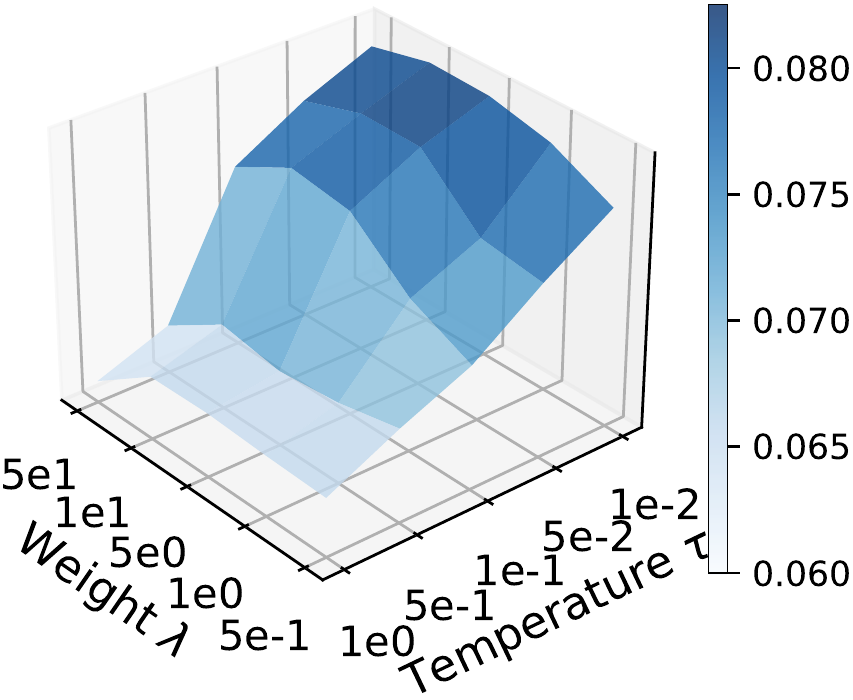}
        \label{fig:hyper3d_pred}
    }
    \subfigure[Embed. Distillation]{
        \includegraphics[width=0.3\columnwidth]{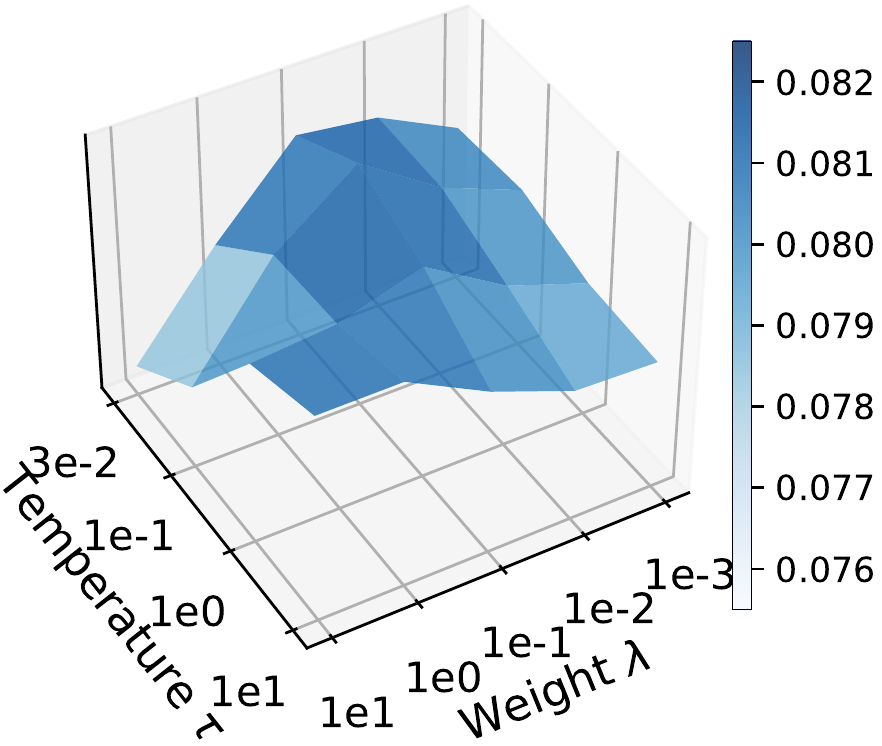}
        \label{fig:hyper3d_embed}
    }
    \subfigure[Contrastive Reg.]{
        \includegraphics[width=0.3\columnwidth]{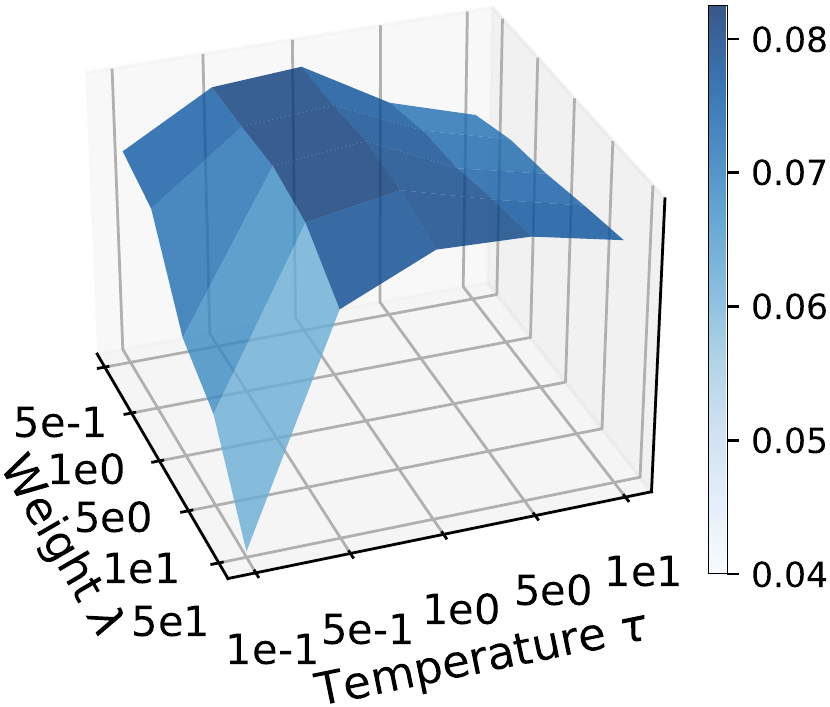}
    }
    \vspace{-0.17in}
    \caption{Impact of weights and temperature in different learning objectives on Yelp, in terms of \emph{Recall@20}.}
    \vspace{-0.2in}
    \label{fig:hyper3d}
    \Description{A three-D figure showing the composite effect of the weight and the temperature coefficient on the performance, for the prediction-level distillation, the embedding-level distillation, and the contrastive regularization.}
\end{figure}
\subsection{Hyperparameter Study (RQ4)}
In this section, we examine the influence of different hyperparameters on the performance of \model. The effect of loss weights $\lambda_1, \lambda_2, \lambda_3$ are shown in Figure~\ref{fig:hyper2d}. The composite effect of loss weights and corresponding temperatures $\tau_1, \tau_2, \tau_3$ are shown in Figure~\ref{fig:hyper3d}. The effect of the size $|\mathcal{T}_1|$ for the prediction-level distillation is shown in Table~\ref{tab:batch_hyper}. Our observations are as follows:
\begin{itemize}[leftmargin=*]
    \item \textbf{Strength of Prediction-Level Distillation}. $\lambda_1, \tau_1$: This weight $\lambda_1$ and temperature $\tau_1$ jointly control the strength of the prediction-level KD $\lambda_1$. We first study the influence of $\lambda_1$ in Figure~\ref{fig:hyper2d} with $\tau_1$ fixed. When $\lambda_1$ is small, not enough knowledge is distilled to the student model which results in deficient performance. When $\lambda_1$ is too large, $\mathcal{L}_1$ cover up the optimization of main loss and yield degraded performance. Additionally, Figure~\ref{fig:hyper3d_pred} shows the positive effect of applying smaller $\tau_1$ to produce larger gradients.
    
    \item \textbf{Strength of Embedding-Level Distillation}. $\lambda_2, \tau_2$: The parameters control the strength of \model\ in restricting the embeddings in MLP to be close to embeddings in GNN. From Figure~\ref{fig:hyper3d_embed} it can be observed that $\lambda_2$ and $\tau_2$ jointly adjust the strength of embedding KD to have modest influence on optimization, to prevent from insufficient knowledge distillation and too-strict embedding regularization. Either large weight with low temperature or small weight with high temperature causes performance decay.
    
    \item \textbf{Strength of Contrastive Regularization} $\lambda_3, \tau_3$: These parameters determine the strength of push-away regularization for preventing over-smoothing. The results show that either too small weight $\lambda_3$ or too high temperature $\tau_3$ causes insufficient regularization and produces over-smoothed embeddings. Meanwhile, strong regularization may damage the modeling of node-wise affinity, and also yields worse performance.
    
    \item \textbf{Per-Batch Number of Samples to Distill} $|\mathcal{T}_1|$: This hyperparameter determines how many instances are sampled to conduct the prediction-level distillation in each training step. According to the results in Table~\ref{tab:batch_hyper}, increasing batch size brings better KD performance until the performance saturates. We ascribe this to the effect that larger batch size filters low-frequency noise in predictions made by the teacher model in \model.
\end{itemize}

\begin{table}[t]
    \caption{Investigation on the impact of batch size in the prediction-oriented distillation of the proposed \model.}
    \vspace{-0.15in}
    \centering
    \footnotesize
    \begin{tabular}{c|c|cccccc}
        \hline
        \multirow{2}{*}{Data} & \multirow{2}{*}{Metric} & \multicolumn{6}{c}{Batch Size $|\mathcal{T}_1|$ in Prediction-Level Distillation}\\
        \cline{3-8}
        & & $1e3$ & $5e3$ & $1e4$ & $5e4$ & $1e5$ & $5e5$\\
        \hline
        \hline
        \multirow{2}{*}{Gowalla} & Recall & 0.2208 & 0.2361 & 0.2399 & 0.2420 & 0.2434 & 0.2448\\
        & NDCG & 0.1441 & 0.1530 & 0.1554 & 0.1577 & 0.1592 & 0.1597\\
        \hline
        \multirow{2}{*}{Yelp} & Recall & 0.0443 & 0.0730 & 0.0773 & 0.0802 & 0.0823 & 0.0822\\
        & NDCG & 0.0210 & 0.0372 & 0.0392 & 0.0407 & 0.0414 & 0.0414\\
        \hline
    \end{tabular}
    \vspace{-0.2in}
    \label{tab:batch_hyper}
    \Description{A table recording the performance change of \model\ with respect to the }
\end{table}

\vspace{-0.1in}
\subsection{Over-Smoothing Investigation (RQ5)}
To investigate whether our graph-less \model\ framework is able to mitigate the over-smoothing effect in graph-structured relation learning for CF, we compare representative baselines and our \model\ model on the Mean Average Distance (MAD) values~\cite{chen2020measuring} over embeddings for the most popular users and items. The evaluation results are shown in Table~\ref{tab:mad}. Our \model\ has higher MAD values on both user and item embeddings for Gowalla and Yelp data, in comparison to not only GCN model GCCF, but also state-of-the-art SSL frameworks. It can be concluded that our \model\ framework better addresses the over-smoothing issue, by learning more uniform-distributed embeddings for users and items, to better characterize their unique interaction patterns. This should be attributed to the MLP-based inference framework, and the contrastive regularization that adaptively alleviates over-smoothing signals.

\begin{table}[t]
    \caption{Investigation on the ability to address the over-smoothing effect on Gowalla and Yelp data in terms of MAD.}
    \vspace{-0.15in}
    \centering
    \footnotesize
    \begin{tabular}{c|c|cccccc}
        \hline
        \multicolumn{2}{c|}{Data} & GCCF & LightGCN & SGL & NCL & HCCF & \emph{\model}\\
        \hline
        \hline
        \multirow{2}{*}{Gowalla} & User & 0.8276 & 0.8203 & 0.8412 & 0.8088 & 0.8394 & \textbf{0.8576}\\
        & Item & 0.7579 & 0.7614 & 0.7702 & 0.8169 & 0.7905 & \textbf{0.8335}\\
        \hline
        \multirow{2}{*}{Yelp} & User & 0.9226 & 0.9610 & 0.9755 & 0.9640 & 0.9749 & \textbf{0.9819}\\
        & Item & 0.6288 & 0.7095 & 0.7191 & 0.6953 & 0.6246 & \textbf{0.7662}\\
        \hline
    \end{tabular}
    \vspace{-0.1in}
    \label{tab:mad}
    \Description{A table presenting the evaluated MAD value of \model\ and baselines. The MAD value of \model\ is higher.}
\end{table}
\section{Related Work}
\label{sec:relate}
\noindent \textbf{Graph-based Collaborative Filtering}
Inspired by the success of GNNs, a lot of research works have designed various graph neural architectures to build collaborative recommender systems~\cite{wu2020graph, gao2021graphrec}. For example, to model user-item interactions graph, many efforts have been devoted to developing powerful GNN models for message passing, \eg~NGCF~\cite{wang2019neural}, {STGCN}~\cite{zhang2019star} and {GCMC}~\cite{berg2017graph}. GCCF~\cite{chen2020revisiting} and LightGCN~\cite{he2020lightgcn} enrich GNNs in CF by simplifying the GCN architecture.
To increase model scalability and prevent over-smoothing in making recommendations, our \model\ abandons graph encoders in the inference model, and conducts soft embedding smoothing by distilling useful knowledge from the GNN-based teacher model. \\\vspace{-0.12in}

\noindent\textbf{Self-Supervised Learning (SSL) for Recommendation}.
To tackle the challenge of noise and sparsity in recommendation systems, recent research has explored various types of SSL techniques for data augmentation~\cite{yu2021self,wei2022contrastive,xia2022self,chen2023heterogeneous}. For instance, some studies, such as SGL~\cite{wu2021self}, introduce random perturbation to generate additional views for CL. Other approaches, such as HCCF~\cite{xia2022hypergraph} and NCL~\cite{lin2022improving}, incorporate global views to produce semantically related pairs for CL. While these SSL methods have shown promise in addressing the issues caused by noisy and sparse data, they often heavily rely on graph neural networks (GNNs) to generate embeddings. This can result in an over-smoothing effect, limiting the overall representation ability of the recommendation framework. \\\vspace{-0.12in}




\noindent\textbf{Knowledge Distillation for Recommendation}.
Knowledge distillation aims to transfer knowledge from a complex and well-trained teacher model to a simpler student model~\cite{zhanggraph}. It utilizes the predictions of the teacher model to generate informative soft targets for the student model to learn from. In the context of recommender systems, knowledge distillation has been used to develop simpler yet effective models~\cite{kang2020rrd, lee2019collaborative, tang2018ranking, xia2022device}. As examples, Tang~\etal~\cite{tang2018ranking} proposes a method to leverage knowledge distillation for ranking tasks in recommender systems. Xia~\etal~\cite{xia2022device} develop highly-efficient models for on-device recommendations with effective knowledge transferring. Unlike the works that primarily focus on model reduction, our proposed approach, \model, aims to address the over-smoothing issue in state-of-the-art GNN-based CF models. By distilling unbiased signals from GNNs to simple multilayer perceptrons (MLPs), we can reduce the over-smoothing effect and enhance the model representation ability.

\section{Conclusion}
\label{sec:conclusoin}

In this paper, we propose a contrastive knowledge distillation model which adaptively transfers knowledge from the GNN-based teacher model to a small feed-forward network, significantly improving the efficiency and robustness of recommender models. Our designed adaptive contrastive regularization generate unbiased self-supervision signals to alleviate the over-smoothing and noise effects commonly exist in recommender systems. Our comprehensive experiments demonstrate the effectiveness of our method in improving recommendation accuracy and achieving better efficiency when compared to state-of-the-art learning techniques. 


\clearpage
\bibliographystyle{ACM-Reference-Format}
\balance
\bibliography{full_refs}

\clearpage
\appendix \section{Appendix}
\balance
\label{sec:appendix}

\subsection{Learning Algorithm of \model}
\label{sec:learn_alg}
The parameter learning for our \model\ is elaborated in Algorithm~\ref{alg:learn_alg}
\begin{algorithm}[h]
	\caption{Learning Process of \model}
	\label{alg:learn_alg}
	\LinesNumbered
	\KwIn{User-item interaction matrix $\textbf{A}$, loss weights and temperature factors $\lambda_1, \lambda_2, \lambda_3, \lambda_4, \lambda^{(t)}, \tau_1, \tau_2, \tau_3$, learning rate $\eta$, maximum training epochs $E$, number of graph iterations $L$, number of MLP layers $L'$.}
	\KwOut{Trained embeddings $\bar{\textbf{H}}^{(s)}$ and MLP parameters $\textbf{W}$.}
	Initialize model parameters $\bar{\textbf{H}}^{(s)}, \bar{\textbf{H}}^{(t)}, \textbf{W}$\\
	Train the GCN teacher model for well-trained $\bar{\textbf{H}}^{(t)}$ (Eq~\ref{eq:teacher_bpr})\\
	\For{$e=1$ to $E$}{
	    \For{mini-batch $\mathcal{T}_2$ drawn from $\mathcal{E}$} {
    	    Sample a batch of triplet $\mathcal{T}_1$\\
    	    Calculate preference difference $z_{i,j,k}^{(s)}, z_{i,j,k}^{(t)}$ for samples in $\mathcal{T}_1$ (Eq~\ref{eq:difference})\\
    	    Compute loss $\mathcal{L}_1$ for prediction-level KD (Eq~\ref{eq:l1})\\
    	    Calculate loss $\mathcal{L}_2$ for embedding-level KD (Eq~\ref{eq:l2})\\
    	    Calculate the adjustment factor $\omega_i, \omega_j$ for users and items in $\mathcal{T}_2$ (Eq~\ref{eq:omega})\\
    	    Compute loss $\mathcal{L}_3$ for contrastive regularization (Eq~\ref{eq:l3})\\
    	    Calculate $\mathcal{L}_\text{rec}$ for recommendation task\\
    	    Calculate $\mathcal{L}_4$ for weight-decay regularization\\
    	    Calculate overall loss $\mathcal{L}^{(s)}$ for the student (Eq~\ref{eq:overall_loss})\\
    	    \For{each parameter ${\theta}$ in $\{\bar{\textbf{H}}^{(s)},\textbf{W}\}$}{
                ${\theta} = {\theta} - \eta\cdot {\partial \mathcal{L}^{(s)}}/{\partial{\theta}}$;\\
            }
	    }
	}
    \Return all parameters $\bar{\textbf{H}}^{(s)}, \textbf{W}$
    \Description{The algorithm for the learning process of \model.}
\end{algorithm}
\vspace{-0.1in}
\subsection{Implementation Details}
\label{sec:implement}
For fair comparison, we present the hyperparameter settings for implementing the proposed \model\ framework and the baseline methods.
Specifically, our \model\ is implemented with PyTorch, using Adam optimizer and Xavier initializer with default parameters. Training batch size is set as $|\mathcal{T}_1|=100000, |\mathcal{T}_2|=4096$. The dimensionality of embedding vectors is set as $32$. The number of MLP layers is selected from $\{1, 2,3\}$. The number of graph iterations for the teacher model is selected from $\{2, 4, 6\}$. The loss weights $\lambda_1, \lambda_2, \lambda_3$ are tuned from $\{10, 3, 1, 0.3, 0.1, 0.03, 0.01\}$, and the weights $\lambda_4, \lambda^{(t)}$ for weight-decay regularization are tuned from $\{1e^{-3},$ $1e^{-4},$ $1e^{-5},$ $1e^{-6},$ $1e^{-7},$ $1e^{-8}, 0\}$. The temperatures $\tau_1, \tau_2, \tau_3$ are chosen from $\{10, 3, 1, 0.3, 0.1, 0.03, 0.01\}$. Parameter $\varepsilon$ for contrastive regularization adjustment is set as $0.2$.


For the baseline methods, we apply the same Adam optimization algorithm, Xavier parameter initializer, and batch size of 4096 as our \model. The hidden dimensionality for all baselines is also set as 32. Hyperparameters that are shared by baseline methods and our \model, are tuned in the same range as above. Such hyperparameters include the number of GNN layers, the weight for weight-decay regularizer. Specifically, for NCL, HCCF, SGL, SLRec, the weight for supplementary tasks are tuned from $\{1e^{-k}, 3e^{-k}|-1\leq k\leq 6\}$. The temperature hyperparameters are tuned from $\{1e^{-k}, 3e^{-k}|2\leq k\leq -1\}$. For NCL, which conducts K-Means clustering every $n$ epochs, we tune $n$ from $\{1, 2, 3, 4, 5\}$. For baseline methods that employs random message dropout (\eg, LightGCN, SGL), the dropout rate is tuned from $\{0.1, 0.2, 0.3, 0.5, 0.8, 0.9\}$. For models that were trained for rating predictions in the original paper (\eg, AutoR, ST-GCN), we train these methods using pair-wise BPR loss for implicit feedback.
For NCF, we adopt the NeuMF version which combines MLPs with Generalized MF.

\subsection{Ablation Study}
\begin{figure}[t]
    \centering
    \includegraphics[width=0.43\columnwidth]{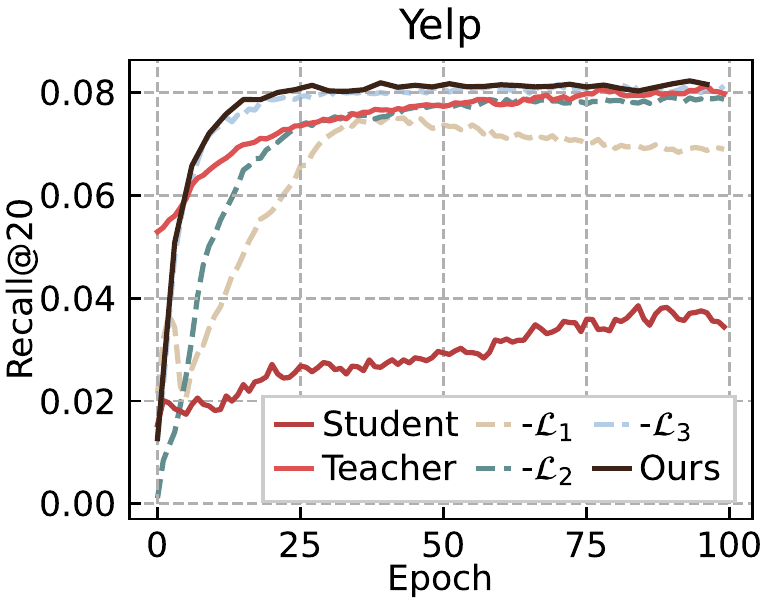}\quad
    \includegraphics[width=0.43\columnwidth]{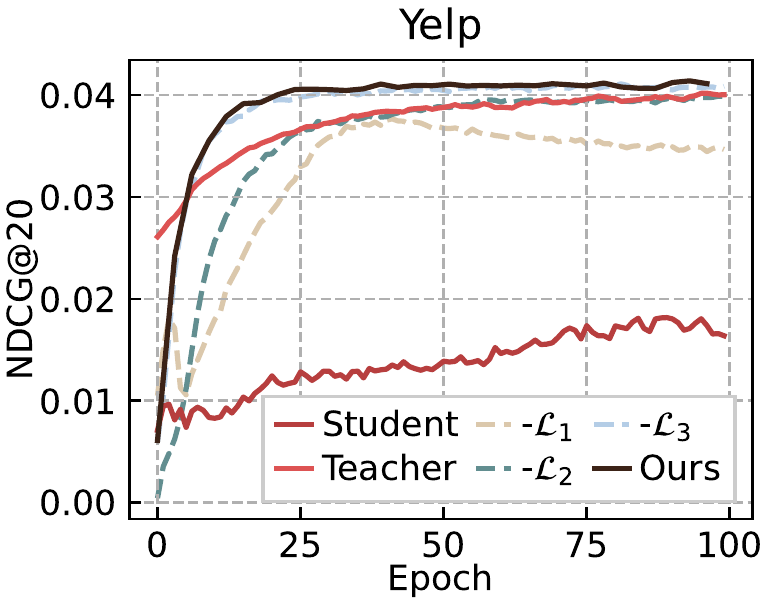}
    \includegraphics[width=0.43\columnwidth]{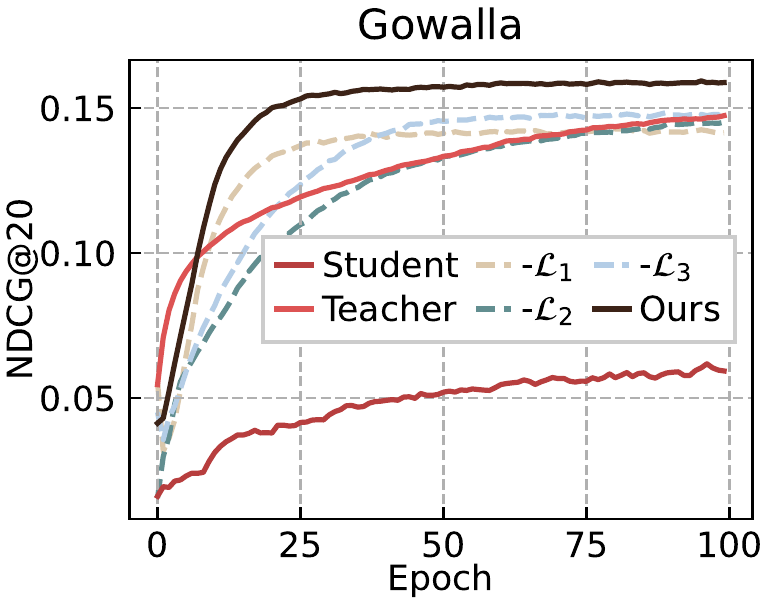}\quad
    \includegraphics[width=0.43\columnwidth]{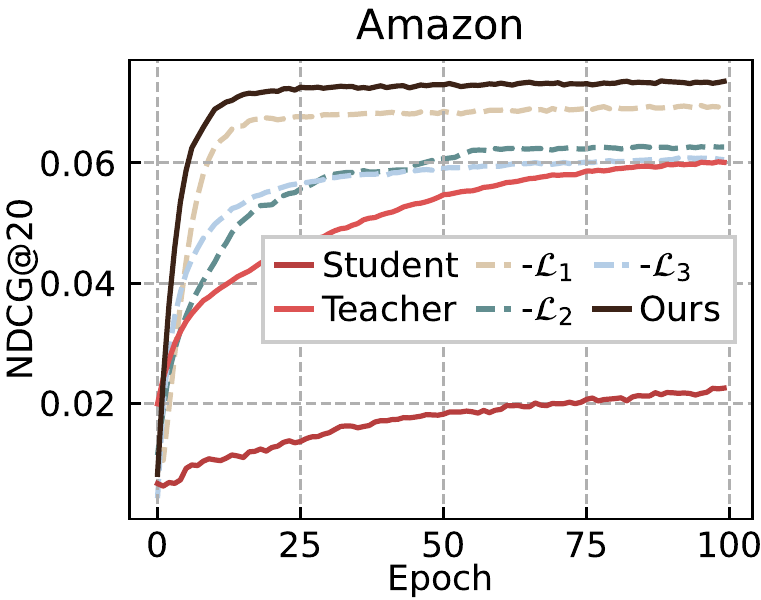}
    \vspace{-0.15in}
    \caption{Test performance in each epoch for ablated models on three experimental datasets in terms of Recall and NDCG.}
    \vspace{-0.15in}
    \label{fig:more_ablation_lines}
    \Description{A line figure showing the performance with respect to epochs for \model\ and some representative baselines. The figure shows that \model\ converges faster while training.}
\end{figure}

We show more results of ablation study in Figure~\ref{fig:more_ablation_lines}, including the Recall@20 and NDCG@20 results on Yelp data, and NDCG@20 results on Gowalla and Amazon data. We can observe that the dual-level knowledge distillation schema and the adaptive contrastive regularization in our \model\ framework significantly improves the performance of the simple MLP model, to even surpass the performance of the GCN-based teacher model. From the results on Yelp data, it can be observed that removing the prediction-level KD causes severe over-fitting. This strongly validates the importance of distilling from the predictions made by the teacher model. Removing the embedding-level distillation, also causes significant performance drop and prominently lower learning efficiency on Yelp data. In comparison, the CL regularization contributes less to the performance of \model\ on Yelp data, which is due to its smaller interaction set that makes it less likely to over-smooth embeddings.


\subsection{Visualization for Embeddings Distribution}
\begin{figure}[t]
    \centering
    \subfigure[NCL]{
        \includegraphics[width=0.3\columnwidth]{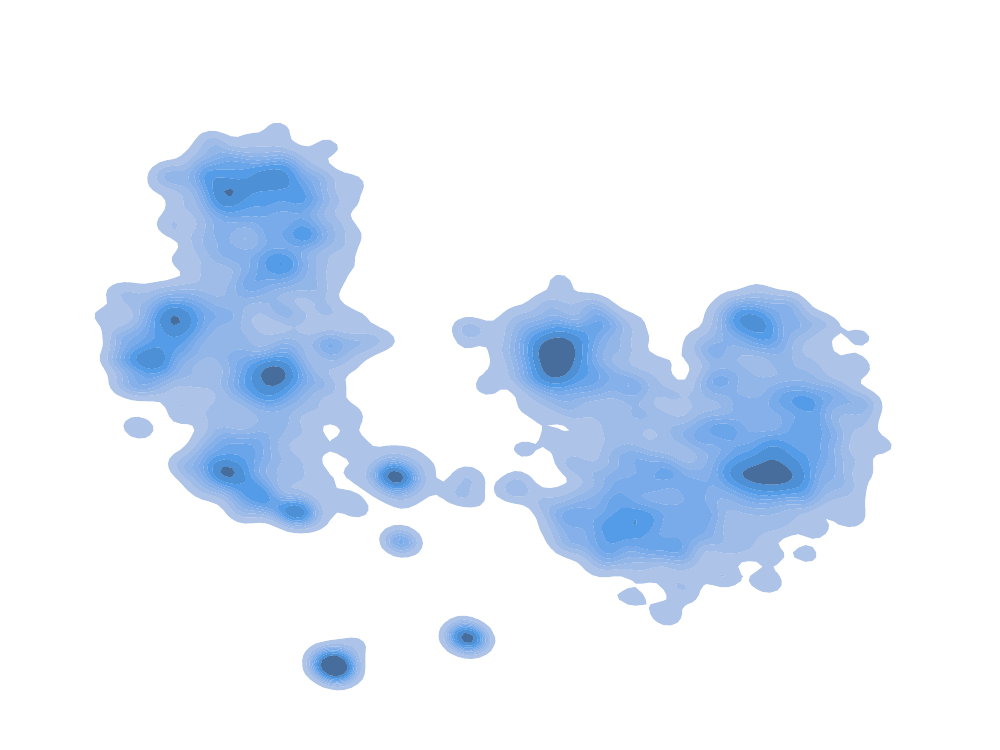}
    }
    \subfigure[HCCF]{
        \includegraphics[width=0.3\columnwidth]{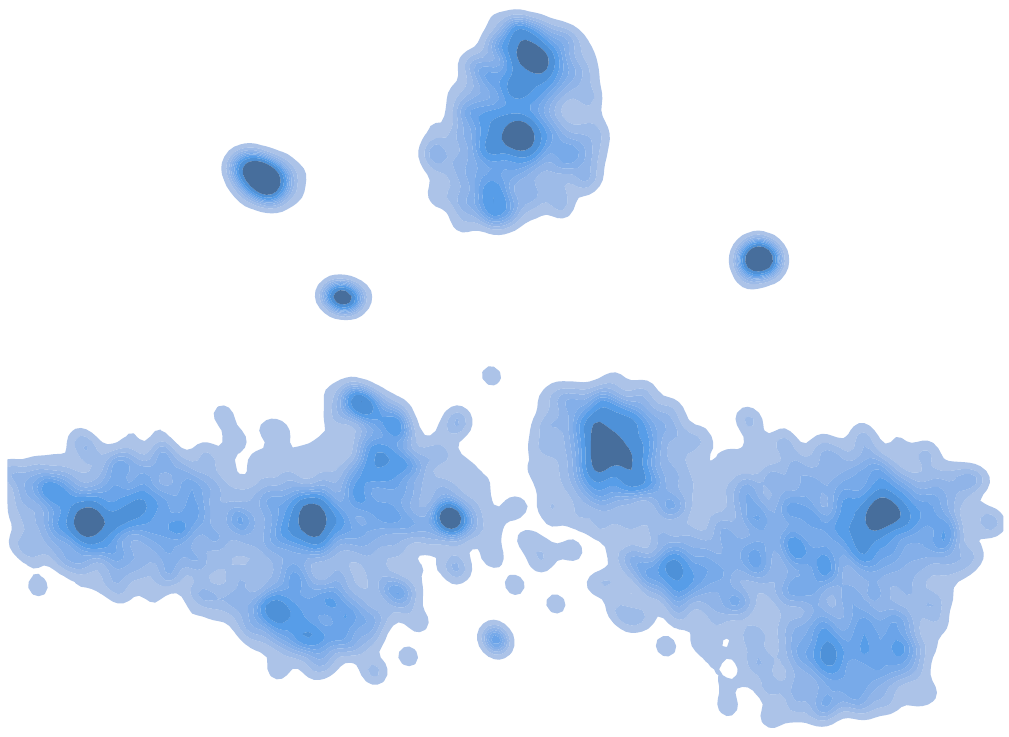}
    }
    \subfigure[\model]{
        \includegraphics[width=0.3\columnwidth]{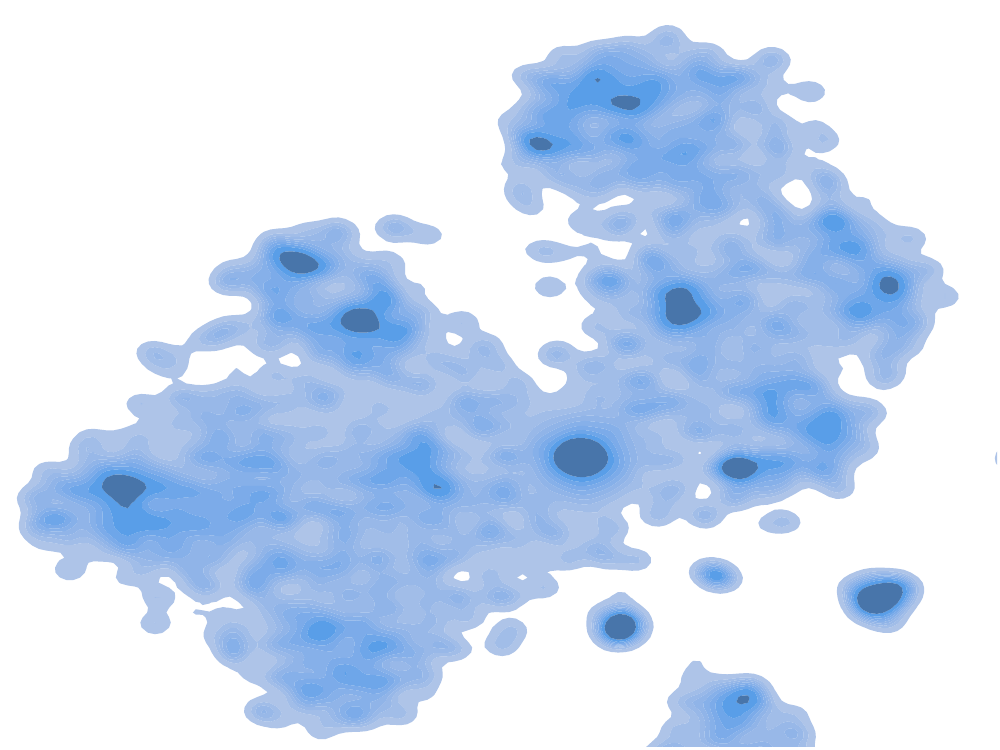}
    }
    \vspace{-0.17in}
    \caption{KDE visualization for distribution of embeddings learned by NCL, HCCF and the proposed \model.}
    \label{fig:more_embeds_dist}
    \vspace{-0.08in}
    \Description{A figure showing the visualized embedding distribution learned by HCCF, NCL and the proposed \model, where the embedding given by \model\ spreads in a wider range.}
\end{figure}

We show more visualization results for the embedding distribution \wrt~NCL in Figure~\ref{fig:more_embeds_dist}. The visualization is done by first compressing the learned embeddings into a 2-d space using t-SNE dimension reduction. Then the scatter plot is smoothed using Gaussian kernel density estimation (KDE) to estimate the distribution of the embeddings. As shown by Figure~\ref{fig:intro_dist} and Figure~\ref{fig:more_embeds_dist}, our \model\ learns to allocates users into a bigger sub-space. In contrast, the baseline methods rely on iterative graph information propagation, which over-smooths the node embeddings to be too similar. From the visualization for the baselines, we can observe that the GNN frameworks over-smooth the user embeddings too much, such that users are split into several prominent subspaces disconnected to each other. This greatly hinders the CF models from learning relations between users from different subspaces.

\subsection{Hyperparameter Study}
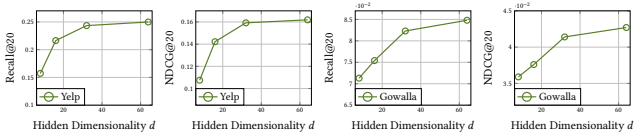
\begin{figure}[t]
    \centering
    \begin{adjustbox}{width=\columnwidth}
    \begin{tikzpicture}
\begin{axis}[
    width=2.8in,
    height=2.4in,
    xlabel={Hidden Dimensionality $d$},
    ylabel= {Recall@20},
    xmin=6,xmax=66,
    ymin=0.1,ymax=0.27,
    legend columns=1,
    legend cell align=right,
    grid=both,
    every axis plot/.append style={ultra thick},
    every tick label/.append style={scale=1.1},
    label style={scale=1.8},
    legend style={
        nodes={scale=1.5, transform shape},
        legend image post style={scale=1.5},
        },
    legend style={at={(0,0)},anchor=south west},
    every axis plot post/.append style={
        every mark/.append style={scale=2}
    }
]
\addplot[color={rgb:red,120;green,171;blue,54}, mark=o, mark options={solid}]
table[x=para, y=yelp_hr] {latFactor.txt};
\legend{\large Yelp};
\end{axis}
\end{tikzpicture}

\begin{tikzpicture}
\begin{axis}[
    width=2.8in,
    height=2.4in,
    xlabel={Hidden Dimensionality $d$},
    ylabel= {NDCG@20},
    xmin=6,xmax=66,
    ymin=0.085,ymax=0.17,
    legend columns=1,
    legend cell align=right,
    grid=both,
    every axis plot/.append style={ultra thick},
    every tick label/.append style={scale=1.1},
    label style={scale=1.8},
    legend style={
        nodes={scale=1.5, transform shape},
        legend image post style={scale=1.5},
        },
    legend style={at={(0,0)},anchor=south west},
    every axis plot post/.append style={
        every mark/.append style={scale=2}
    }
]
\addplot[color={rgb:red,120;green,171;blue,54}, mark=o, mark options={solid}]
table[x=para, y=yelp_ndcg] {latFactor.txt};
\legend{\large Yelp};
\end{axis}
\end{tikzpicture}

\begin{tikzpicture}
\begin{axis}[
    width=2.8in,
    height=2.4in,
    xlabel={Hidden Dimensionality $d$},
    ylabel= {Recall@20},
    xmin=6,xmax=66,
    ymin=0.065,ymax=0.087,
    legend columns=1,
    legend cell align=right,
    grid=both,
    every axis plot/.append style={ultra thick},
    every tick label/.append style={scale=1.1},
    label style={scale=1.8},
    legend style={
        nodes={scale=1.5, transform shape},
        legend image post style={scale=1.5},
        },
    legend style={at={(0,0)},anchor=south west},
    every axis plot post/.append style={
        every mark/.append style={scale=2}
    }
]
\addplot[color={rgb:red,120;green,171;blue,54}, mark=o, mark options={solid}]
table[x=para, y=gowalla_hr] {latFactor.txt};
\legend{\large Gowalla};
\end{axis}
\end{tikzpicture}

\begin{tikzpicture}
\begin{axis}[
    width=2.8in,
    height=2.4in,
    xlabel={Hidden Dimensionality $d$},
    ylabel= {NDCG@20},
    xmin=6,xmax=66,
    ymin=0.032,ymax=0.045,
    legend columns=1,
    legend cell align=right,
    grid=both,
    every axis plot/.append style={ultra thick},
    every tick label/.append style={scale=1.1},
    label style={scale=1.8},
    legend style={
        nodes={scale=1.5, transform shape},
        legend image post style={scale=1.5},
        },
    legend style={at={(0,0)},anchor=south west},
    every axis plot post/.append style={
        every mark/.append style={scale=2}
    }
]
\addplot[color={rgb:red,120;green,171;blue,54}, mark=o, mark options={solid}]
table[x=para, y=gowalla_ndcg] {latFactor.txt};
\legend{\large Gowalla};
\end{axis}
\end{tikzpicture}
    \end{adjustbox}
    \vspace{-0.25in}
    \caption{Hyperparameter study for hidden dimensionality of \model\ in terms of Recall and NDCG on Yelp and Gowalla.}
    \vspace{-0.1in}
    \label{fig:hyperparam_embed}
    \Description{A line figure presenting the performance change of \model\ with respect to the hidden dimensionality.}
\end{figure}

We further investigate the influence of hidden dimensionality in our \model\ for the model performance. Specifically, we first train GCN-based teacher models with different hidden dimensionality (8, 16, 32, 64), and then distill the teacher model to a MLP-based student model with the same embedding size. As shown by results in Figure~\ref{fig:hyperparam_embed}, the performance shows a typical under-fitting to over-fitting curve \wrt~the hyperparameter $d$ on different datasets. After $d$ reaches the default embedding size 32, the performance increases slightly on Yelp dataset. Instead, the performance still prominently grows when $d$ increases from 32 to 64 on Gowalla data. This could be attributed to the larger scale of interaction records and the lower sparsity degree of Gowalla data.

\subsection{Theoretical Analysis}

\subsubsection{\bf Detailed Complexity Analysis}
\label{sec:complexity_analysis}
The complexity analysis is to answer the following two questions: i) How do GCNs compare to MLPs in efficiency? ii) What is the overhead of our KD paradigm?
In each training step, GNN-based CF methods must conduct whole-graph information propagation for the embedding process. This takes $\mathcal{O}(|\mathcal{E}|\times L\times d)$ complexity for our lightweight GCN. The prediction phase of our GCN takes $\mathcal{O}(|\mathcal{T}_\text{bpr}|\times d)$ for computing dot-product. In comparison, the embedding process of MLP is not in graph-level but focus on one embedding vector at once. It costs $\mathcal{O}(|\mathcal{T}_2|\times L'\times d^2)$ where $|\mathcal{T}_2|=|\mathcal{T}_\text{bpr}|\ll|\mathcal{E}|/d$. It also requires $\mathcal{O}( |\mathcal{T}_\text{bpr}| \times d)$ computational cost to predict in each training batch. 


Our prediction-level KD $\mathcal{L}_1$ requires $\mathcal{O}(|\mathcal{T}_1|\times d)$ cost. The $\mathcal{L}_2$ KD takes $\mathcal{O}(|\mathcal{T}_2|\times d)$ for the numerators, and $\mathcal{O}(|\mathcal{T}_2|\times J\times d)$ for the denominators. Similar to the second term for $\mathcal{L}_2$, the contrastive regularization $\mathcal{L}_3$ costs $\mathcal{O}(|\mathcal{T}_2|\times (I+J)\times d)$ computations. In conclusion, the KD of our \model\ has the total time complexity of $\mathcal{O}(|\mathcal{T}_2|\times (I+J)\times d)$, which is comparable to the state-of-the-art CF methods (\eg, self-supervised methods SGL~\cite{wu2021self}, NCL~\cite{lin2022improving}). Note that although the training process has the same complexity, our \model\ conducts inference with simple MLPs which is much more efficient as discussed above.

\subsubsection{\bf Derivations for High-Order Smoothing}
\label{sec:embed_analysis}
In this section, we present details for the derivations related to Section~\ref{sec:highorder_smoothing}. To begin with, we show the high-order smoothing effect of the GCN teacher in the perspective of gradients, which yield the results in Eq~\ref{eq:gcn_gradient}. Specifically, the gradients that maximize the similarity between  $\bar{\textbf{h}}_i^{(t)}$ and $\bar{\textbf{h}}_j^{(t)}$, given by the loss $\mathcal{L}^{(t)}$ is as follows:
\begin{align}
    \frac{\partial \mathcal{L}_{i,j}}{\partial \bar{\textbf{h}}_i}
    &=  -\sum_{u_i,v_j,v_k} \frac{\partial\log \text{sigm}(z_{i,j,k})}{\partial \bar{\textbf{h}}_i}
    = -\sum_{u_i,v_j,v_k} \sigma\frac{\partial z_{i,j,k}}{\partial \bar{\textbf{h}}_i}\nonumber\\
    &= -\sum_{u_i,v_j,v_k}\sigma\frac{\partial\textbf{h}_i^\top\textbf{h}_j}{\partial\bar{\textbf{h}}_i}
    =-\sum_{v_k}\sigma \sum_{n_{i'},n_{j'}}\sum_{\mathcal{P}_{i,i'}^L, \mathcal{P}_{j,j'}^L} \frac{\partial \textbf{h}_{i'}^\top\textbf{h}_{j'}}{\partial\bar{\textbf{h}}_i}\nonumber\\
    &=-\sum_{v_k}\sigma \sum_{n_{i'},n_{j'}}\sum_{\mathcal{P}_{i,i'}^L, \mathcal{P}_{j,j'}^L} \prod_{(n_{a},n_{b})\in\mathcal{P}_{i,i'}^L}\frac{1}{\sqrt{d_a d_b}}\nonumber\\
    &\prod_{(n_a,n_b)\in\mathcal{P}_{j,j'}^L}\frac{1}{\sqrt{d_ad_b}}
    \frac{\partial\bar{\textbf{h}}_i^\top\bar{\textbf{h}}_j}{\partial\bar{\textbf{h}}_i}\nonumber\\
    &=\sum_{v_k}- \sigma \cdot
    \Big(\sum_{\mathcal{P}_{i,j}^{2L}} \prod_{(n_a,n_b)\in\mathcal{P}_{i,j}^{2L}} \frac{1}{\sqrt{d_a d_b}}\Big)\cdot
    \frac{\partial\bar{\textbf{h}}_i^{\top}\bar{\textbf{h}}_j^{}}{\partial \bar{\textbf{h}}_i}
\end{align}
\noindent where $\sigma$ denotes $1-\text{sigm}(z_{i,j,k})$. For simplicity, we omit the $(t)$ superscript. As $\mathcal{L}_{i,j}$ refers to the pull-close terms, $-\textbf{h}_i^\top\textbf{h}_k$ is omitted. Next, we show the details of derivations that obtain Eq~\ref{eq:pd_gradient} as follows:
\begin{align}
    {\mathcal{L}_1}
    &= \sum_{u_i,v_j,v_k}-\bar{z}_{i,j,k}^{(t)} \cdot \log \bar{\textbf{z}}_{i,j,k}^{(s)} + (\bar{z}_{i,j,k}^{(t)} - 1) \cdot \log(1 - \bar{\textbf{z}}_{i,j,k}^{(s)})\nonumber\\
    &= \sum_{u_i,v_j,v_k} \bar{z}_{i,j,k}^{(t)} \cdot \log\frac{1-\bar{z}_{i,j,k}^{(s)} }{\bar{z}_{i,j,k}^{(s)}} - \log(1 - \bar{z}_{i,j,k}^{(s)})\nonumber\\
    &=\sum_{u_i,v_j,v_k} -\bar{z}_{i,j,k}^{(t)} {z}_{i,j,k}^{(s)} /\tau_1+\log(1+\exp(z_{i,j,k}^{(s)}/\tau_1))\nonumber\\
    \frac{\partial\mathcal{L}_1}{\partial\textbf{h}_i^{(s)}}
    &= \sum_{u_i,v_j,v_k} \frac{\text{sigm}(z_{i,j,k}^{(s)}/\tau_1)}{\tau_1} \frac{\partial\textbf{h}_i^{(s)\top}\textbf{h}_j^{(s)}}{\partial\textbf{h}_i^{(s)}} - \frac{1}{\tau_1}\bar{z}_{i,j,k}^{(t)} \frac{\partial\textbf{h}_i^{(s)\top}\textbf{h}_j^{(s)}}{\partial\textbf{h}_i^{(s)}} \nonumber\\
    &=\sum_{u_i,v_j,v_k} -\frac{1}{\tau_1} \cdot (\bar{z}_{i,j,k}^{(t)} - \bar{z}_{i,j,k}^{(s)}) \cdot \frac{\partial\textbf{h}_i^{(s)\top} \textbf{h}_j^{(s)}}{\partial\textbf{h}_i^{(s)}}
\end{align}
From the derivation above, we can observe that GCN conduct high-order embedding smoothing using the cumulative product of node degrees as weights. This manner is restricted by the graph structures and may be affected by noisy edges. Instead, our developed \model\ uses knowledge distillation to perform adaptive high-order smoothing for any user-item pair $u_i,v_j$, using the teacher model's predictions as guidance during the model compression process. This allows the lightweight student model to effectively learn from the teacher's knowledge and make accurate predictions.


\end{document}